\documentclass[fleqn,10pt]{wlscirep}
\usepackage[utf8]{inputenc}
\usepackage[T1]{fontenc}
\usepackage{epsfig}
\usepackage{epstopdf}
\usepackage{lineno}
\usepackage{amsmath}
\usepackage{setspace} 
\usepackage{tikz}
\usepackage{helvet}
\usepackage{caption}
\usepackage{amssymb}
\usepackage{bbding}
\usepackage{epstopdf}
\usepackage{xcolor}
\usepackage{bm}
\usepackage{morefloats}
\usepackage{anyfontsize}
\usepackage{siunitx}
\usepackage{makecell} 
\newcolumntype{C}[1]{>{\centering\arraybackslash}m{#1}} 

\newcommand{\ra}[1]{\renewcommand{\arraystretch}{#1}}
\newcommand{\co}[1]{\setlength\tabcolsep{#1}}

\newcommand{\beginextended}{%
	\setcounter{table}{0}
	\setcounter{figure}{0}
}
\newcommand{\extendeddatafigure}{%
	\captionsetup[figure]{name=Extended Data Fig.}
}
\newcommand{\extendeddatatable}{%
	\captionsetup[table]{name=Extended Data Table}
}

\newcommand{\beginsupplement}{%
	\setcounter{table}{0}
	\setcounter{figure}{0}
}
\newcommand{\supplementfigure}{%
	\captionsetup[figure]{name=Supplementary Fig.}
}

\newcommand{\supplementtable}{%
	\captionsetup[table]{name=Supplementary Table}
}

\title{Bridging the behavior-neural gap: A multimodal AI reveals the brain's geometry of emotion more accurately than human self-reports}

\author[1,3,$\dag$]{Changde Du}
\author[1,2,$\dag$]{Yizhuo Lu}
\author[1,3,$\dag$]{Zhongyu Huang}
\author[1,3]{Yi Sun}
\author[4]{Zisen Zhou}
\author[4]{Shaozheng Qin}
\author[1,2,3,*]{Huiguang He}
\affil[1]{State Key Laboratory of Brain Cognition and Brain-inspired Intelligence Technology, Institute of Automation, Chinese Academy of Sciences, Beijing, China}
\affil[2]{School of Future Technology, University of Chinese Academy of Sciences, Beijing, China}
\affil[3]{School of Artificial Intelligence, University of Chinese Academy of Sciences, Beijing, China}
\affil[4]{State Key Laboratory of Cognitive Neuroscience and Learning, Beijing Normal University, Beijing, China}

\affil[$\dag$]{These authors contributed equally.}
\affil[*]{corresponding author: Huiguang He (huiguang.he@ia.ac.cn)}

\begin{abstract}
	The ability to represent emotion plays a significant role in human cognition and social interaction, yet the high-dimensional geometry of this affective space and its neural underpinnings remain debated. A key challenge, the ‘behavior-neural gap,’ is the limited ability of human self-reports to predict brain activity. Here we test the hypothesis that this gap arises from the constraints of traditional rating scales and that large-scale similarity judgments can more faithfully capture the brain's affective geometry. Using AI models as ‘cognitive agents,’ we collected millions of triplet odd-one-out judgments from a multimodal large language model (MLLM) and a language-only model (LLM) in response to 2,180 emotionally evocative videos. We found that the emergent 30-dimensional embeddings from these models are highly interpretable and organize emotion primarily along categorical lines, yet in a blended fashion that incorporates dimensional properties. Most remarkably, the MLLM's representation predicted neural activity in human emotion-processing networks with the highest accuracy, outperforming not only the LLM but also, counterintuitively, representations derived directly from human behavioral ratings. This result supports our primary hypothesis and suggests that sensory grounding—learning from rich visual data—is critical for developing a truly neurally-aligned conceptual framework for emotion. Our findings provide compelling evidence that MLLMs can autonomously develop rich, neurally-aligned affective representations, offering a powerful paradigm to bridge the gap between subjective experience and its neural substrates. Project page: \href{https://reedonepeck.github.io/ai-emotion.github.io/}{ai-emotion.github.io}

\end{abstract}

\begin{document}
	
	\flushbottom
	\maketitle
	
	\section*{Introduction}
	How is emotion encoded in the brain and abstracted into a computable cognitive representation? This question remains a central puzzle in understanding human intelligence. For decades, the scientific debate on the structure of emotion has been dominated by two competing views: one posits that emotions are comprised of a set of discrete, universal categories like joy and fear \cite{ekman1992argument, prinz2004emotions, panksepp2004affective, ledoux2012rethinking}, while the other contends that they are constructed from a few continuous dimensions, such as valence and arousal \cite{russell1980circumplex, barrett2006emotions, russell2003core, lindquist2012brain, barrett2017emotions}. 
	This long-standing dichotomy, however, may fundamentally overlook the true nature of affective representation---a high-dimensional, blended geometry that transcends simple categorical or dimensional frameworks \cite{cowen2021semantic,keltner2023semantic}. More recent theories, such as the theory of constructed emotion, posit that emotions are not monolithic entities but are constructed from more fundamental psychological and neural components in the moment \cite{barrett2017emotions}. This perspective offers a theoretical framework for understanding modern neuroscientific evidence, which suggests that emotion is not supported by discrete, emotion-specific modules but emerges from dynamic interactions within large-scale brain networks that implement more fundamental psychological processes \cite{ lindquist2016brain,kragel2016decoding, zhang2025neurofunctional, saarimaki2018discrete, phan2002functional}.  Yet, despite our growing understanding of this neural complexity \cite{vceko2022common, bo2024systems,reddan2025neural,zhou2021distributed,liu2024neural,gan2024neurofunctional,zhang2021role}, models of emotion built from human self-reports (e.g., rating scales) show limited power in predicting this neural activity \cite{horikawa2020neural,koide2020distinct,giordano2021representational,du2023topographic,lettieri2024dissecting,saarimaki2025cerebral}, creating an intractable `behavior-neural gap.' 
	
	Traditional rating scales, by forcing subjective experience into a fixed set of predefined labels or dimensions \cite{cowen2017self,cowen2020music}, may inadvertently constrain and distort the very phenomena they seek to measure. Participants may struggle to map their nuanced feelings onto discrete numerical scales, and the choice of labels itself introduces experimenter bias. An alternative paradigm, based on large-scale similarity judgments (e.g., millions of triplet odd-one-out choices), circumvents these limitations. By relying on simple, qualitative decisions, this approach minimizes experimenter priors and allows the underlying representational structure to emerge organically from the data itself \cite{hebart2020revealing, contier2024distributed, mahner2025dimensions}. 
	This leads to a critical, untested hypothesis: the behavior-neural gap in affective science may arise from the constraints and distortions imposed by traditional rating scales. A comprehensive map of emotion, constructed from millions of similarity judgments, might more faithfully reflect the brain's native affective geometry. This hypothesis, however, has remained purely theoretical, as it is practically hard for human participants to complete behavioral experiments of such a massive scale.
	
	Could we, then, find a valid agent for human emotional judgment? In recent years, large language models (LLMs) have demonstrated impressive, often human-level, performance on a wide range of complex cognitive tasks \cite{binz2025foundation, du2025human, luo2025large, xu2025large, cui2025large, lu2025cultural}. Crucially, recent studies have already confirmed their considerable sensitivity to affective content \cite{li2024language, schlegel2025large, rubin2025comparing, huang2024apathetic, bojic2025comparing, wang2023emotional, sabour2024emobench, li2023good}. This capability is understood to arise not from phenomenal experience, but from a profound, implicit knowledge of the statistical regularities of human expression, learned from vast corpora of human language and behavior. Their `decisions' in a similarity judgment task can thus be conceptualized as a high-throughput, data-driven reflection of the conventional structure of human emotion. Based on this, we pioneer the `cognitive agent' paradigm: a framework that employs LLMs as indefatigable cognitive agents to perform tasks previously hard for humans, and ultimately, validating this surrogacy against the gold standard of alignment with real brain activity.

	Furthermore, this powerful paradigm also provides an unprecedented opportunity to address a central debate in cognitive science: the role of sensory grounding in the formation of abstract concepts \cite{barsalou2008grounded,barsalou2010grounded,lecun2023large,chemero2023llms,dove2024symbol,xu2025large}. For non-sensorimotor concepts such as emotion, the necessity of grounding is far from settled \cite{niedenthal2007embodying,barrett2008embodiment,winkielman2018dynamic}. Indeed, recent work suggests that language-only models, even without direct sensory connection to the world, can successfully represent non-sensorimotor features of concepts, including emotion \cite{xu2025large}. This leads to a more nuanced and critical question, which we term the `neural alignment hypothesis for sensory grounding': while language alone may be sufficient to form a coherent semantic space for emotion, is sensory grounding nonetheless essential for developing a representational geometry that is truly aligned with the neural architecture of human affective processing? To directly test this hypothesis, our study systematically compares two types of artificial intelligence (AI) models with distinct cognitive architectures: a language-only model (LLM), which learns from abstract textual symbols, and a vision-language multimodal LLM (MLLM), which learns from both visual and textual data, a process more akin to the visual sensory experience of humans.
	
\begin{figure}[!h]
		\centering
        \includegraphics[width=17.4cm]{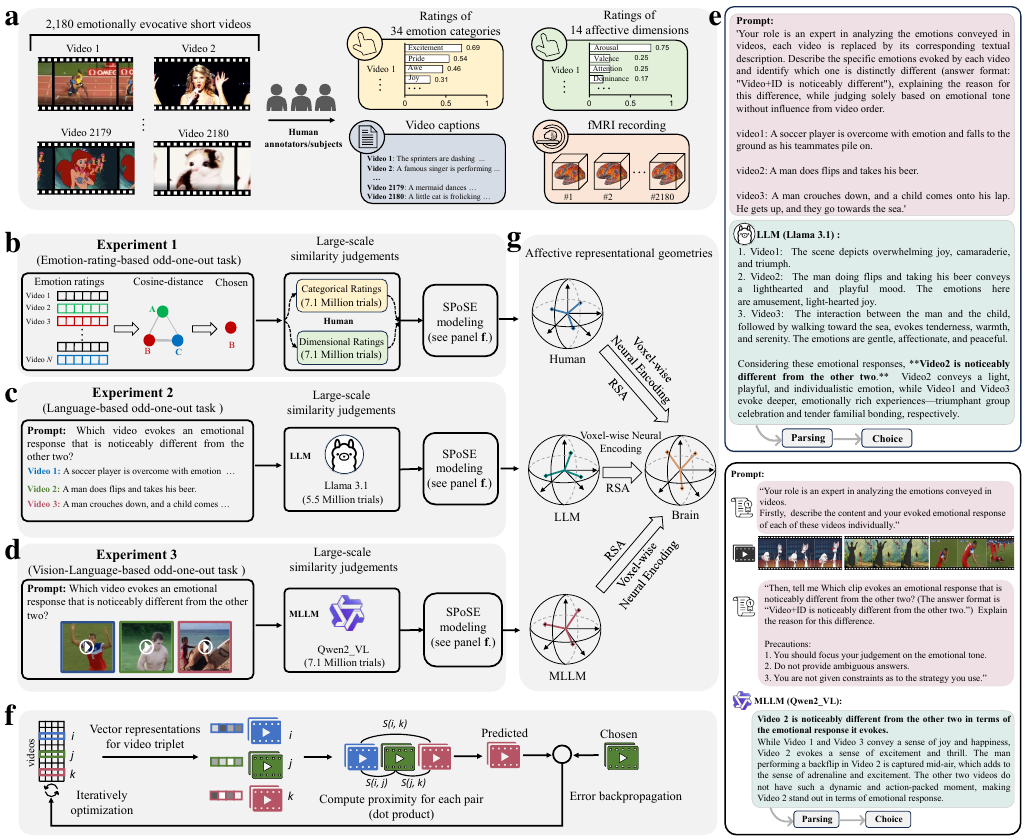}   
		\caption{\textbf{Overview of the experimental and analytical pipeline.} \textbf{a}, The study utilized a database of 2,180 emotionally evocative videos with rich, pre-existing annotations, including human ratings on discrete emotion categories and continuous affective dimensions, detailed textual descriptions, and corresponding fMRI data from viewers. \textbf{b-d}, Affective embeddings were derived for four systems—human categorical ratings, human dimensional ratings, LLM, and MLLM—using a triplet odd-one-out behavioral paradigm. Human similarity judgments were simulated based on the cosine similarity of their prior ratings, while models performed the task directly. \textbf{e}, Example prompts and responses for the LLM and MLLM. \textbf{f-g}, Latent embeddings were learned from over 7.1 million triplet judgments using SPoSE (\textbf{f}), and the resulting representational spaces were compared to each other and to neural data (\textbf{g}).}
		\label{fig:figure1} 
	\end{figure}

	To test these hypotheses, we introduce a framework based on large-scale machine-behavioral experiments. By prompting an LLM and an MLLM to make millions of similarity judgments on 2,180 emotionally evocative videos (or their descriptions), we map their emergent affective representational spaces and systematically compare them to human behavioral ratings and fMRI data (Fig. \ref{fig:figure1}). The results yield a striking discovery: the MLLM's emergent representation predicts neural activity in the human brain with significantly greater accuracy than models derived from human behavioral ratings themselves. This superior neural alignment provides the strong empirical support for our central hypothesis—that a paradigm based on large-scale similarity judgments can indeed capture the brain's native affective geometry more faithfully than traditional rating scales. Further analysis of this neurally-aligned representation reveals that it is highly interpretable; its structure, while primarily organized along categorical lines, artfully blends features of continuous dimensions, forming a hybrid representation that is highly consistent with modern Semantic Space Theory of emotion \cite{cowen2021semantic,keltner2023semantic} and thus computationally reconciles the long-standing `category' versus `dimension' debate \cite{barrett2006emotions, ekman1992argument}. Finally, the superior performance of the multimodal model offers compelling computational neuroscience evidence for the theory of sensory grounding, suggesting that a connection to the world through sensory modalities is key to developing a truly neurally-aligned conceptual framework for emotion. This work not only establishes MLLM as a powerful tool for exploring the computational underpinnings of affective cognition but also opens a path toward bridging the gap between subjective experience and its neural substrates.
	
	\section*{Results}
	
	To construct a comprehensive map of the human affective space—a goal previously intractable due to the immense scale of data required—we introduced a novel framework that leverages large AI models as indefatigable `cognitive agents'. Our approach centered on a large-scale dataset of 2,180 emotionally evocative videos, which are richly annotated with human emotion ratings \cite{cowen2017self}, textual descriptions \cite{horikawa2024mind}, and corresponding fMRI activity \cite{horikawa2020neural} (Fig. \ref{fig:figure1}a). We probed the representational geometry of each system using a triplet odd-one-out behavioral paradigm \cite{zheng2019revealing, hebart2020revealing, mahner2025dimensions, du2025human} (Fig. \ref{fig:figure1}b-e), generating over 5.5 million similarity judgments for the LLM (Llama-3.1 \cite{touvron2023llama}, using text captions) and 7.1 million similarity judgments for the MLLM (Qwen2-VL \cite{wang2024qwen2}, using video input). For a human benchmark, we simulated judgments based on prior human ratings across 34 emotion categories and 14 affective dimensions \cite{cowen2017self}, creating two distinct human representational models. From these four sets of behavioral judgments, we used Sparse Positive Similarity Embedding (SPoSE) \cite{zheng2019revealing} to learn a 30-dimensional affective embedding space for each system (Fig. \ref{fig:figure1}f). Finally, we compared the structure of these learned spaces to each other and assessed their alignment with human brain activity using representational similarity analysis (RSA) \cite{kriegeskorte2008representational} and voxel-wise encoding (Fig. \ref{fig:figure1}g).
	
	Before assessing their neurobiological plausibility, we first validated that the learned affective embeddings are both stable and predictive. The 30-dimensional spaces, a dimensionality at which predictive performance saturated, were highly reproducible across independent model initializations (Extended Data Fig. \ref{fig:Dimensional-Validity}a-e). Crucially, these embeddings demonstrated excellent predictive validity: they accurately predicted triplet choices on held-out data, capturing the majority of explainable variance (up to 91.59\% for MLLM), and their global representational structure generalized to unseen stimuli ($r$ up to 0.85 for MLLM; Extended Data Fig. \ref{fig:Dimensional-Validity}f-j). This confirms that our machine-behavioral paradigm yields robust and generalizable affective representations, providing a solid foundation for the primary analyses that follow.
	\begin{figure}[!h]	
		\centering
        \includegraphics[width=17.4cm]{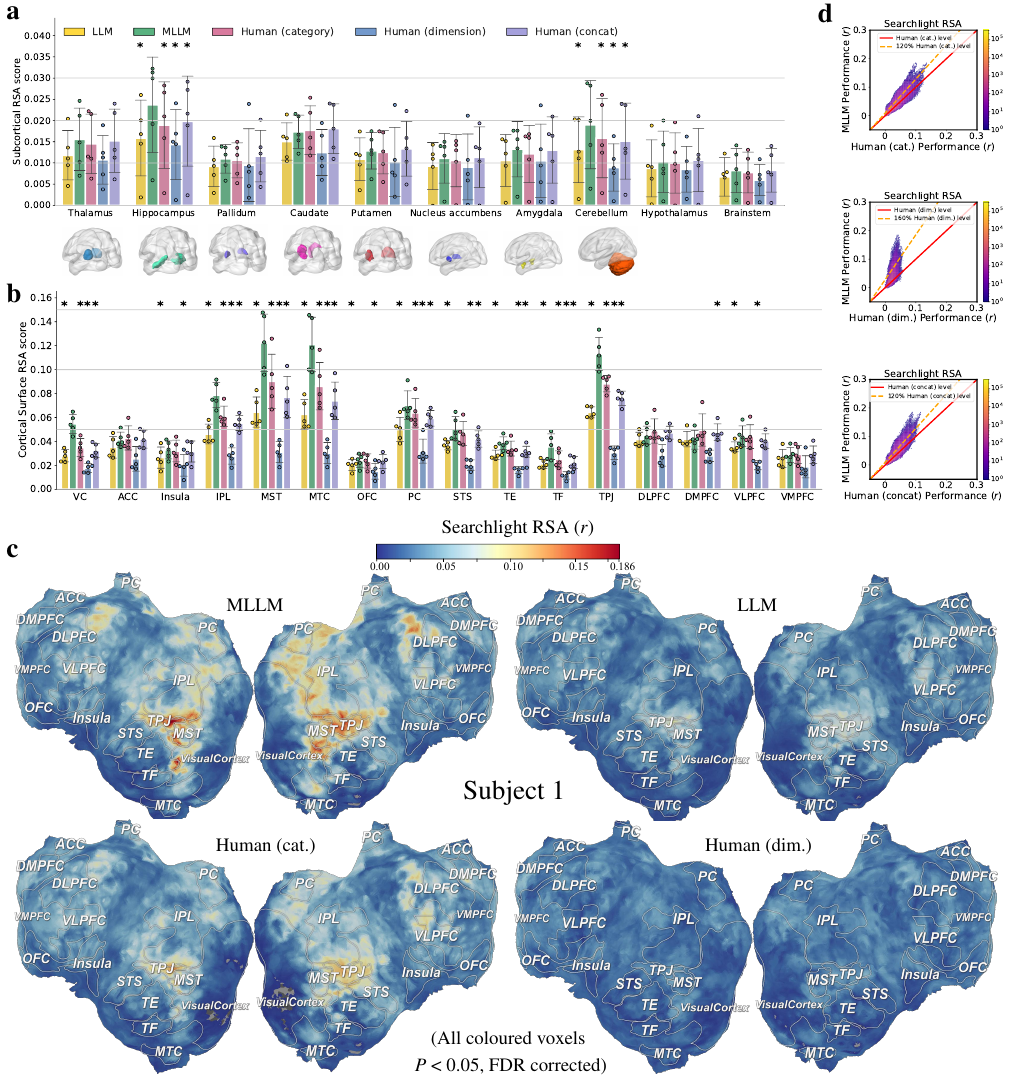} 
		\caption{\textbf{MLLM representations show superior alignment with neural representations of emotion.} \textbf{a, b}, Results of searchlight RSA between model and brain, averaged in subcortical (\textbf{a}) and cortical (\textbf{b}) ROIs (averaged across subjects (N=5)). Dots represent individual subjects; error bars reflect standard deviation (s.d.); all statistics are two-tailed t-tests across subjects, with false discovery rate (FDR) correction; stars indicate significant differences between MLLM and the compared model ($P$ < 0.05). \textbf{c}, Whole-cortex maps of searchlight RSA for a representative subject, illustrating the MLLM's superior performance across distributed emotion-processing networks. All coloured voxels are predicted significantly ($P$ < 0.05, FDR-corrected, two-tailed t-tests). \textbf{d}, Voxel-wise comparison of the MLLM's performance against the human categorical, dimensional and concatenated ratings model, using searchlight RSA. }
		
		\label{fig:compare_with_brain} 
	\end{figure}
	
	\subsection*{Bridging the behavior-neural gap with affective representations derived from MLLM.}
	Our central hypothesis was that a comprehensive affective map, derived from large-scale similarity judgments, would bridge the `behavior-neural gap' by aligning more closely with the brain's native geometry. To directly test this, we compared the representational geometry of each model to that of human brain activity recorded during video viewing. A searchlight RSA \cite{kriegeskorte2008representational} across the brain revealed a two-fold result (Fig. \ref{fig:compare_with_brain}a,b).
	
	First, and most remarkably, the MLLM's representation outperformed the representations derived directly from human behavioral ratings. Voxel-wise encoding analyses in Extended Data Fig.~\ref{fig:RSA_encoding_appendix}a,b also reveal consistent trends favoring the MLLM. To ensure this counterintuitive finding was not an artifact of using separate human rating systems, we performed a critical control analysis by creating a comprehensive `Human-concat' model that combined categorical and dimensional ratings. Even this unified human model was substantially outperformed by the MLLM. This finding is robust, as the SPoSE-derived human embedding used for this comparison is a faithful representation of the original behavioral data, demonstrating a comparable level of neural alignment to the raw self-report ratings themselves (Extended Data Fig.~\ref{fig:spose_vs_raw_comparison}). This result provides the strong empirical support for our central hypothesis: the `behavior-neural gap' in affective science may be a direct consequence of the methodological constraints of traditional rating scales. The machine-behavioral paradigm, by enabling similarity judgments at an unprecedented scale, appears to capture the brain's affective geometry with higher fidelity, allowing the MLLM to converge on a representational scheme that is fundamentally more neurally-aligned.
	
	Second, the MLLM's representation showed significantly higher alignment with neural patterns than its language-only LLM counterpart. This finding provides powerful computational evidence for the `sensory grounding' hypothesis, suggesting that a direct connection to the world through rich, dynamic visual data—rather than abstract linguistic symbols—is critical for an AI model to develop a truly neurally-aligned affective framework.
	
	The MLLM's superior neural alignment was particularly evident in a distributed network of regions critical for emotion and social cognition. Whole-cortex searchlight maps (Fig. \ref{fig:compare_with_brain}c, Extended Data Fig. \ref{fig:RSA_encoding_appendix}c and Extended Data Figs. \ref{fig:sub2_cortex}-\ref{fig:sub5_cortex}a,b) revealed that the MLLM exhibited a significant advantage in several cortical areas, including the temporoparietal junction (TPJ) \cite{lettieri2019emotionotopy, saxe2013people,decety2007role, schiller2019theta}, the inferior parietal lobule (IPL), high-level visual cortex (VC) \cite{kragel2019emotion, gao2025object}, the medial superior temporal area (MST) \cite{nassi2009parallel, dukelow2001distinguishing, huk2002retinotopy}, and the medial temporal cortex (MTC). Furthermore, this advantage extended to key subcortical and cerebellar structures, including the hippocampus and the cerebellum. Additionally, in Extended Data Fig. \ref{fig:dim_vis}, we projected several affective components (\textit{the dimensions of the learned embedding space, so termed to distinguish them from theoretical affective dimensions like valence}) of MLLM onto the cortical surface. The results reveal that a single affective component is processed concurrently across multiple functionally related brain regions, albeit with varying activation intensities. A voxel-wise comparison across the entire brain further confirmed the MLLM's robust advantage (Fig. \ref{fig:compare_with_brain}d, Extended Data Fig. \ref{fig:RSA_encoding_appendix}d). The vast majority of voxels showed higher alignment with the MLLM than with the best-performing human model, with the MLLM's performance reaching up to 160\% of the human model's performance. This pattern was highly consistent across all five subjects (Extended Data Figs. \ref{fig:sub2_cortex}-\ref{fig:sub5_cortex}c,d), underscoring the robustness of this discovery. 
	\begin{figure}[!h]
		\centering
        \includegraphics[width=15.8cm]{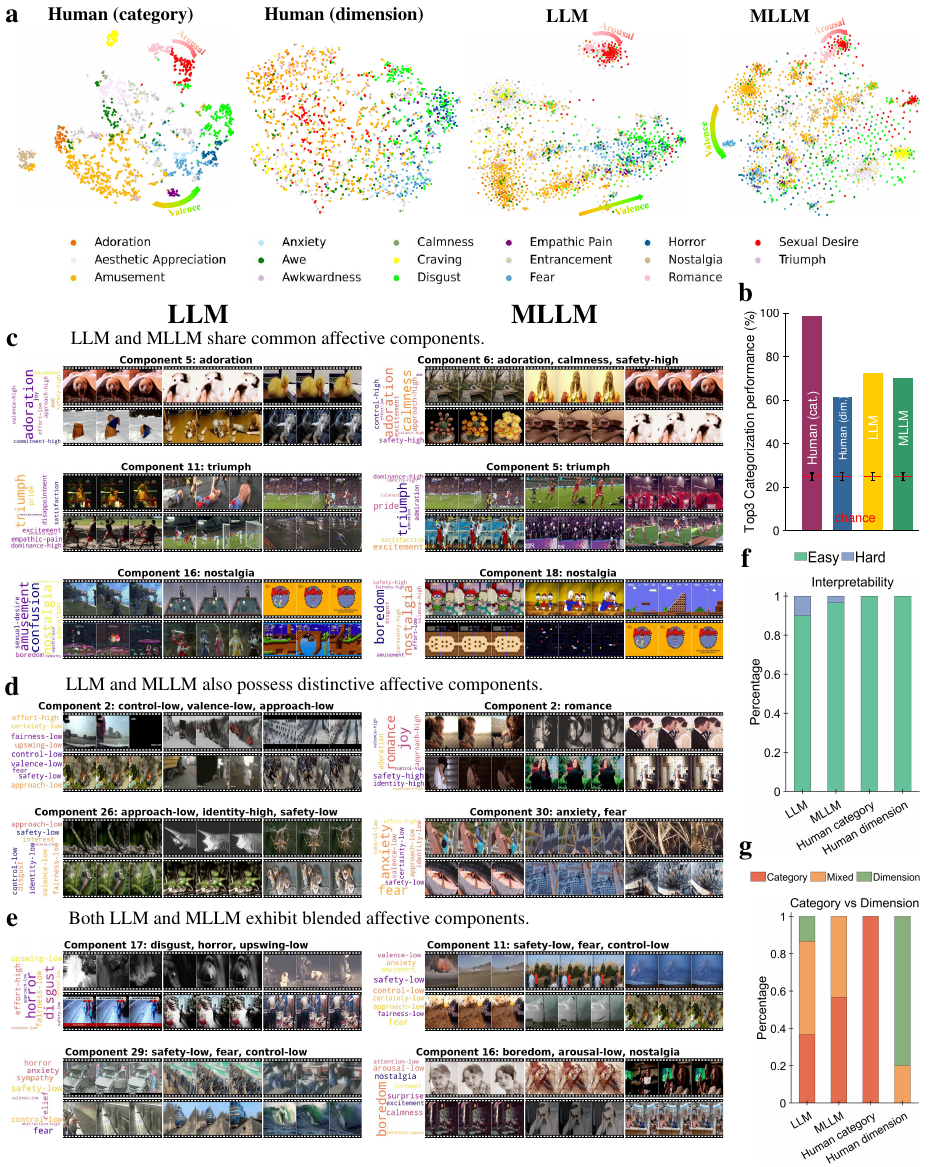} 
		\caption{\textbf{Models develop interpretable, categorical, and blended affective components.} \textbf{a}, t-SNE visualization of the 2,180 stimuli reveals the global structure of the affective spaces. Points are colored by their highest-rated human emotion category, showing spontaneous clustering. \textbf{b}, Top-3 nearest-centroid accuracy, quantifying the categorical structure. \textbf{c-e}, Examples of shared (\textbf{c}), unique (\textbf{d}), and blended (\textbf{e}) affective components, with top-weighted video frames and word clouds (label size $\propto$ correlation with human ratings). \textbf{f}, Proportion of components that were interpretable versus uninterpretable. \textbf{g}, Proportion of interpretable components best described as purely categorical, purely dimensional, or a mix of both.}
		\label{fig:Dimensional-Vis}
	\end{figure}
	
	\subsection*{Models develop interpretable, categorical, and blended affective representations.}
	This remarkable brain alignment, particularly the MLLM's advantage over human-derived models, raises a critical question: What is the underlying structure of this neurally-superior affective space? To understand why this representation is so effective, we dissected its emergent geometry, probing its interpretability and its fundamental organizational principles to see if it conforms to traditional theories or finds a novel synthesis of its own.
	
	To visualize the global structure of the learned embeddings, we projected the 2,180 video stimuli into a two-dimensional space using t-SNE (Fig. \ref{fig:Dimensional-Vis}a). Despite receiving no explicit categorical labels during the triplet judgment task, both LLM and MLLM spontaneously organized their representations into distinct clusters corresponding to human emotion categories (e.g., \textit{amusement}, \textit{fear}, \textit{craving}). These emergent structures mirrored those derived from human categorical ratings and, to a lesser extent, human dimensional ratings, underscoring the foundational role of categorical organization in emotion representation \cite{cowen2017self, horikawa2020neural}. Moreover, the visualizations revealed smooth, psychologically meaningful transitions between clusters along continuous axes, such as increasing arousal linking \textit{romance} to \textit{sexual desire}, and decreasing valence connecting \textit{amusement} to \textit{fear}. These patterns closely align with established models of human affective space \cite{cowen2017self, horikawa2020neural, cowen2021semantic, keltner2023semantic}.
	
	A nearest-centroid classifier confirmed the strong categorical structure within the embeddings (Fig. \ref{fig:Dimensional-Vis}b). While embeddings from human categorical ratings performed best (top-3 accuracy: $\SI{98.58}{\percent}$), both LLM ($\SI{72.43}{\percent}$) and MLLM ($\SI{70.14}{\percent}$) substantially outperformed chance, demonstrating robust categorical separability. To interpret the individual affective components, we correlated each one with human ratings of 34 emotion categories and 14 affective dimensions \cite{cowen2017self} (see Extended Data Table \ref{table:30dim_labels} and Extended Data Figs. \ref{fig: LLM-MLLM_extend1}-\ref{fig: human_extend3}). This revealed that the vast majority of components were highly interpretable across all systems (Fig. \ref{fig:Dimensional-Vis}f; We defined a component as interpretable if the correlation between it and any known label is greater than 0.1).
	
	The models developed both shared and unique affective components, revealing their distinct cognitive architectures. For example, both LLM and MLLM formed a component for \textit{adoration} in response to cute animals and another for \textit{triumph} from victorious moments (Fig.~\ref{fig:Dimensional-Vis}c). However, models' unique components offered compelling evidence for the `sensory grounding' hypothesis (Fig.~\ref{fig:Dimensional-Vis}d). The LLM, constrained to abstract symbols, formed abstract components like \textit{low-control} for scenes of disasters, whereas the MLLM leveraged its direct perceptual access to the visual world to make far more granular distinctions, such as separating \textit{romance} from \textit{sexual desire} and identifying \textit{anxiety} in videos of extreme sports—nuances the LLM missed. This divergence demonstrates how sensory grounding enables a fine-grained, object-centric affective representation, whereas a purely symbolic system must compensate by relying on more abstract conceptual relationships.
	
	Ultimately, the most profound discovery was that both models converged on a hybrid coding scheme that intrinsically blended categorical and dimensional features (Fig.~\ref{fig:Dimensional-Vis}e). For instance, LLM associated scenes of car crashes with both \textit{horror} (a category) and \textit{low-upswing} (a dimension). Across the models, we found that 15 of the LLM's components and 13 of the MLLM's were of this mixed type (Fig. \ref{fig:Dimensional-Vis}g). This organization, which is neither purely categorical nor purely dimensional, strongly aligns with the Semantic Space Theory of human emotion \cite{cowen2021semantic, keltner2023semantic}, suggesting a convergent organizational principle for affective representation in both artificial and biological intelligence.
	
	\begin{figure}[!th]
		\centering
        \includegraphics[width=17.4cm]{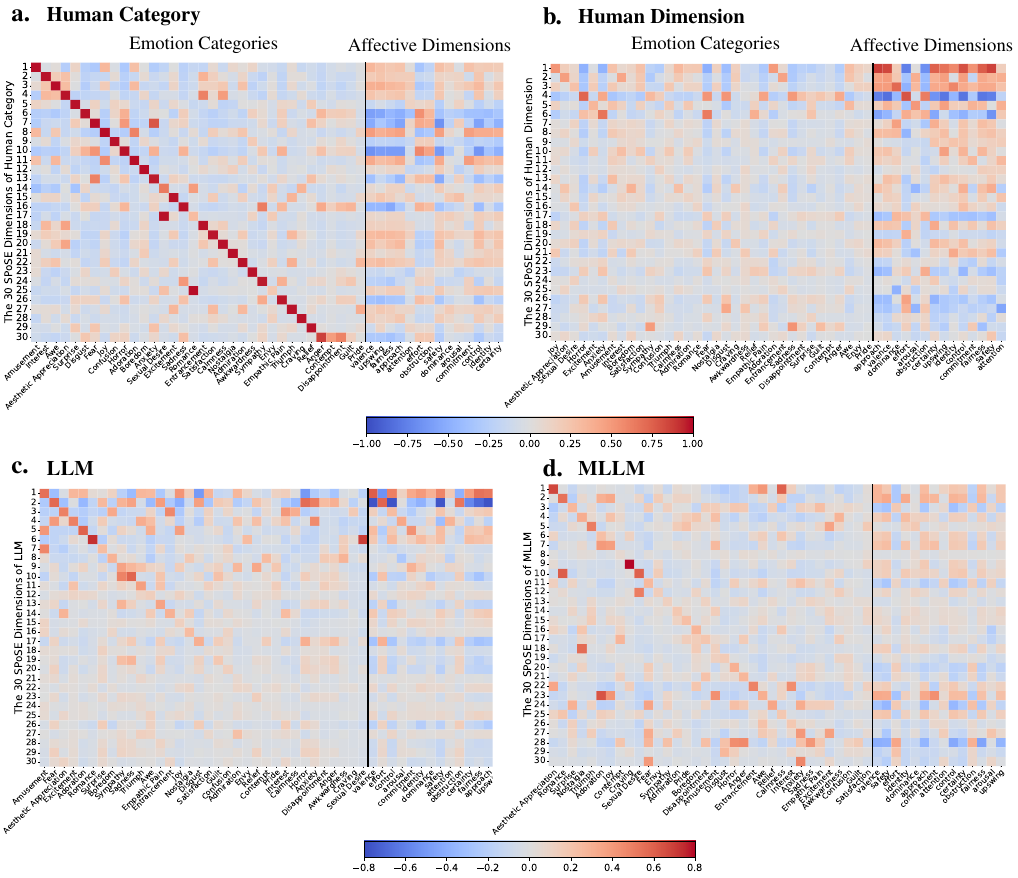}
		\caption{\textbf{Learned affective spaces are primarily organized by emotion categories.} Correlation heatmaps showing the relationship between the 30 learned affective components (y-axes) and the 48 categories/dimensions from human self-reports (x-axes; 34 categories and 14 dimensions from Cowen et al. \cite{cowen2017self}). Each cell represents the PCC between a learned component and a human-rated category or dimension. The strong diagonal patterns observed for the LLM (\textbf{c}) and MLLM (\textbf{d})—mirroring the pattern from human categorical data (\textbf{a}) but not dimensional data (\textbf{b})—indicate that the models' affective spaces are predominantly structured along categorical, rather than dimensional, lines.}
		\label{fig:river_plot}
	\end{figure}
	
	\subsection*{Models computationally reconcile the category-dimension debate with a hybrid coding scheme.}
	
	A central debate in affective science concerns whether emotions are represented as discrete categories \cite{ekman1992argument, prinz2004emotions, panksepp2004affective, ledoux2012rethinking} or as constructions of continuous dimensions \cite{barrett2006emotions, russell2003core, lindquist2012brain, barrett2017emotions}. Our neurally-aligned machine representations provide a unique computational testbed to adjudicate this long-standing dichotomy. 
	Fig.~\ref{fig:river_plot} shows the full cross-correlation matrices between the 30 learned components and the human ratings across 34 emotion categories and 14 affective dimensions \cite{cowen2017self}. As expected, the human-derived embeddings served as unambiguous benchmarks: the `human-category' space showed a clean, one-to-one mapping with discrete categories (Fig.~\ref{fig:river_plot}a, $r>0.95$ for 28 components; Extended Data Table \ref{table:30dim_labels}), while the `human-dimension' space strongly mapped onto continuous dimensions (Fig.~\ref{fig:river_plot}b). 
	
	Both the LLM and MLLM representations revealed a sophisticated synthesis of these two frameworks. On one hand, they showed a powerful allegiance to the \textit{categorical} benchmark, with their learned components forming a strong diagonal pattern analogous to the human-category data (Fig.~\ref{fig:river_plot}c,d). This indicates that a categorical structure serves as the primary organizing backbone of the models' affective space. On the other hand, the representations were not rigidly categorical. Unlike the cleanly separated human-category data, a single model component often showed significant correlations not only with several related emotion categories (e.g., MLLM \textit{component-27} correlating with \textit{anxiety}, \textit{excitement}, and \textit{fear}) but also with one or more continuous affective dimensions. 
	
	This evidence demonstrates that (M)LLMs do not adopt either theoretical framework exclusively but instead forge a hybrid, blended coding scheme. This emergent organization offers a data-driven, computational reconciliation for the long-standing `category' versus `dimension' debate, providing compelling support for the Semantic Space Theory \cite{cowen2021semantic, keltner2023semantic}, which posits that human emotion concepts are simultaneously categorical and graded.
	
	\begin{figure}[!h]
		\centering
        \includegraphics[width=17.4cm]{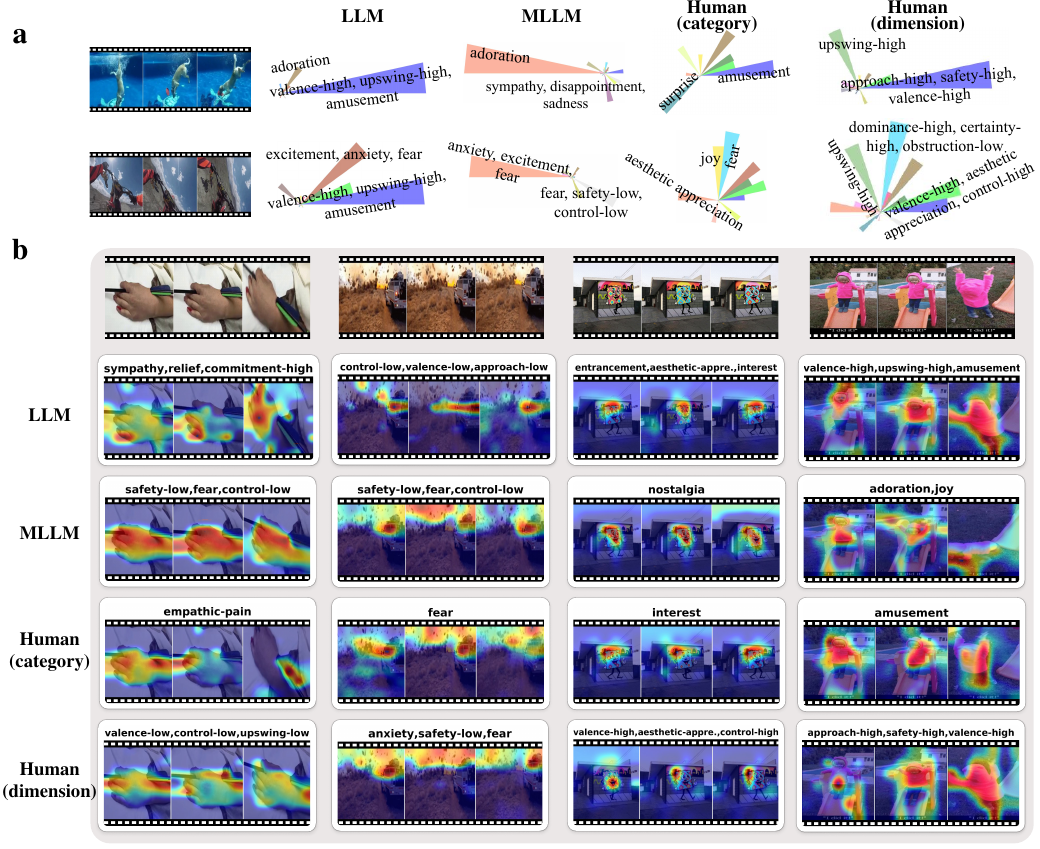}
		\caption{\textbf{Attribution methods reveal the visual drivers of affective components.} \textbf{a}, Illustration of example videos with their affective decomposition. Each colored petal represents a specific affective component, and its length is proportional to that component's weight for the given video. \textbf{b}, Grad-CAM heatmaps identify the specific visual regions that drive responses in selected components. Red indicates a strong positive contribution from that region to the component's activation.}
		\label{fig:Explainable_gradCAM}
	\end{figure}
	
	\subsection*{Learned affective components functionally control visual features.}
	
	Having established the hybrid structure of the affective space, we next sought to probe the functional role of its individual constituents. Decomposing individual videos into their constituent affective components revealed consistent and interpretable affective signatures across all systems. Rose diagrams (Fig. \ref{fig:Explainable_gradCAM}a), where each petal represents a component and its length reflects the component's weight, illustrate this consistency. For instance, a video of a dog playing in a pool reliably elicited the component of \textit{amusement} and \textit{high valence} in both humans and models. Similarly, a first-person perspective of a bicycle skydiving stunt was consistently decomposed into components of \textit{anxiety}, \textit{excitement}, and \textit{fear}.

	To move beyond correlational evidence and investigate the functional role of these components, we developed a two-stage validation framework. First, we used gradient-weighted class activation mapping (Grad-CAM) \cite{selvaraju2017grad} to localize the precise visual features that drove the activation of specific affective components. This attribution analysis revealed that the models' affective components were linked to semantically appropriate visual elements (Fig. \ref{fig:Explainable_gradCAM}b). For example, the MLLM's \textit{fear} component was activated by the sight of a hand pierced by an arrow, with the heatmap concentrated on the injury. The LLM's \textit{valence-low} component was triggered by an exploding vehicle, with attribution focused on the blast regions. In a positive context, the MLLM's \textit{adoration} component was activated by a child on a slide, with the heatmap highlighting the child.
	
	These attribution patterns were notably consistent across the MLLM, LLM, and human-derived representations. Remarkably, despite processing only textual descriptions and having no direct visual input, LLM correctly inferred the likely visual objects responsible for eliciting specific emotions, generating heatmaps nearly indistinguishable from the MLLM's in many cases. However, this result does not suggest that sensory experience is superfluous. Rather, it reveals that human language is itself a vast repository of implicitly encoded sensory knowledge. Through large-scale language training, the LLM has learned to construct a sophisticated conceptual blueprint of emotion, one that robustly links abstract affective states to their most probable and prototypical visual causes \cite{xu2025large, li2024language, schlegel2025large}. The LLM's success, therefore, lies not in replicating perception itself, but in masterfully reverse-engineering a conceptual model of it from the statistical shadow it leaves in language.
	
\begin{figure}[!h]
		\centering
        \includegraphics[width=17.4cm]{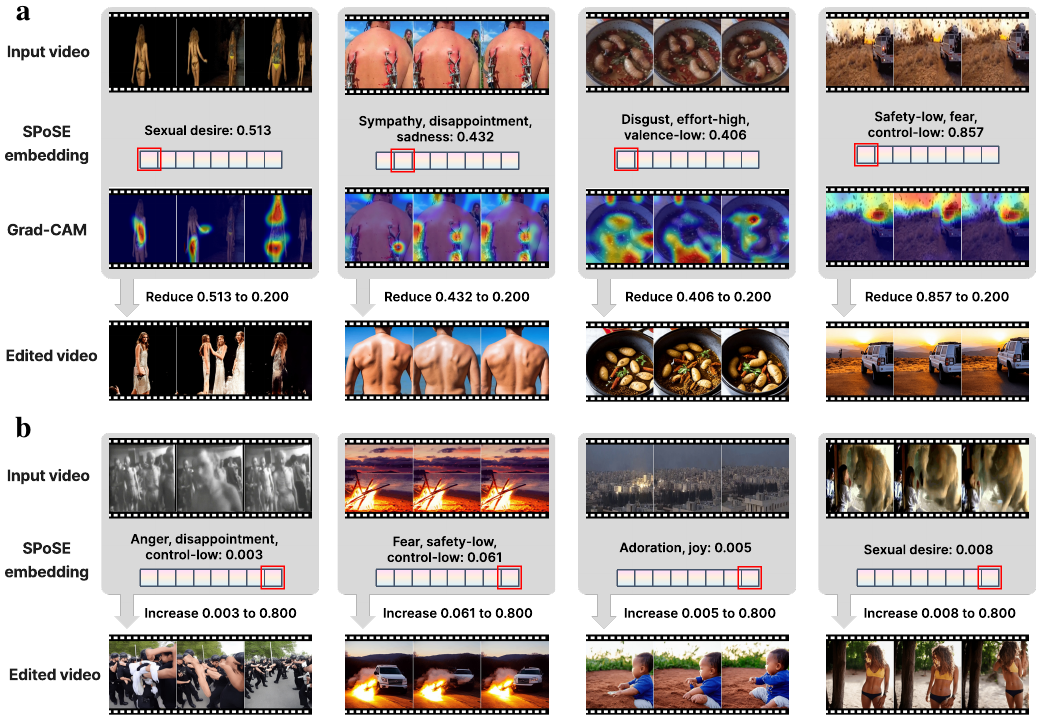}
		\caption{\textbf{Generative editing reveals how learned affective components control video content.} \textbf{a}, Targeted removal of emotional content by decreasing the activation of a specific affective component. Top to bottom: original video frames; the targeted component (marked by red box) and its corresponding label as well as original activation (normalized to [0,1]); the Grad-CAM heatmap identifying the visual drivers; and the edited video frames, where decreasing the component's activation transforms the visual drivers into irrelevant contents. \textbf{b}, Generative elicitation of emotional content by increasing component activations. For example, increasing the \textit{fear} component's activation converts a peaceful campfire scene into a dangerous scene with a vehicle leaping through flames. These manipulations demonstrate that the learned components functionally control the visual elements that evoke specific emotions. For closed-loop validation, see Extended Data Figs. \ref{fig: edit_MLLM_145} and \ref{fig: edit_MLLM_67}.}	
		\label{fig:Editing}
	\end{figure}

	The second stage of our framework, generative editing, provided strong evidence for the functional influence of the learned components. By directly manipulating the activation value of a single component, we could predictably and precisely alter the visual content of the video (Fig. \ref{fig:Editing}). For instance, decreasing the activation of the \textit{sexual desire} component (from 0.513 to 0.2) transformed a runway model's undergarments into a formal dress (Fig. \ref{fig:Editing}a). Similarly, attenuating a \textit{sympathy} component removed the chains from a man's back piercings. Conversely, increasing a component's activation could generate emotionally congruent content. Elevating the \textit{anger, disappointment} component (from 0.003 to 0.8) converted a documentary scene of a concentration camp into footage of a riot, while amplifying the \textit{fear, safety-low} component transformed a peaceful campfire into a dangerous scene with a vehicle leaping through flames (Fig. \ref{fig:Editing}b). Extended Data Figs. \ref{fig: edit_MLLM_145} and \ref{fig: edit_MLLM_67} present closed-loop validation results where MLLM was presented with both original and edited videos for categorical judgment and explanation. The MLLM's altered responses to edited videos confirm successful manipulation of its affective interpretations (Extended Data Figs. \ref{fig: edit_MLLM_145} and \ref{fig: edit_MLLM_67}).  Extended Data Fig. \ref{fig:emo_edit_appendix} presents more editing results. Together, these experiments provide compelling evidence for a functional link: the learned affective components appear to not merely \textit{correlate} with emotional content, but to actively \textit{encode} and \textit{influence} the specific visual features that generate an affective response.

	\begin{figure}[!h]
		\centering
        \includegraphics[width=17.4cm]{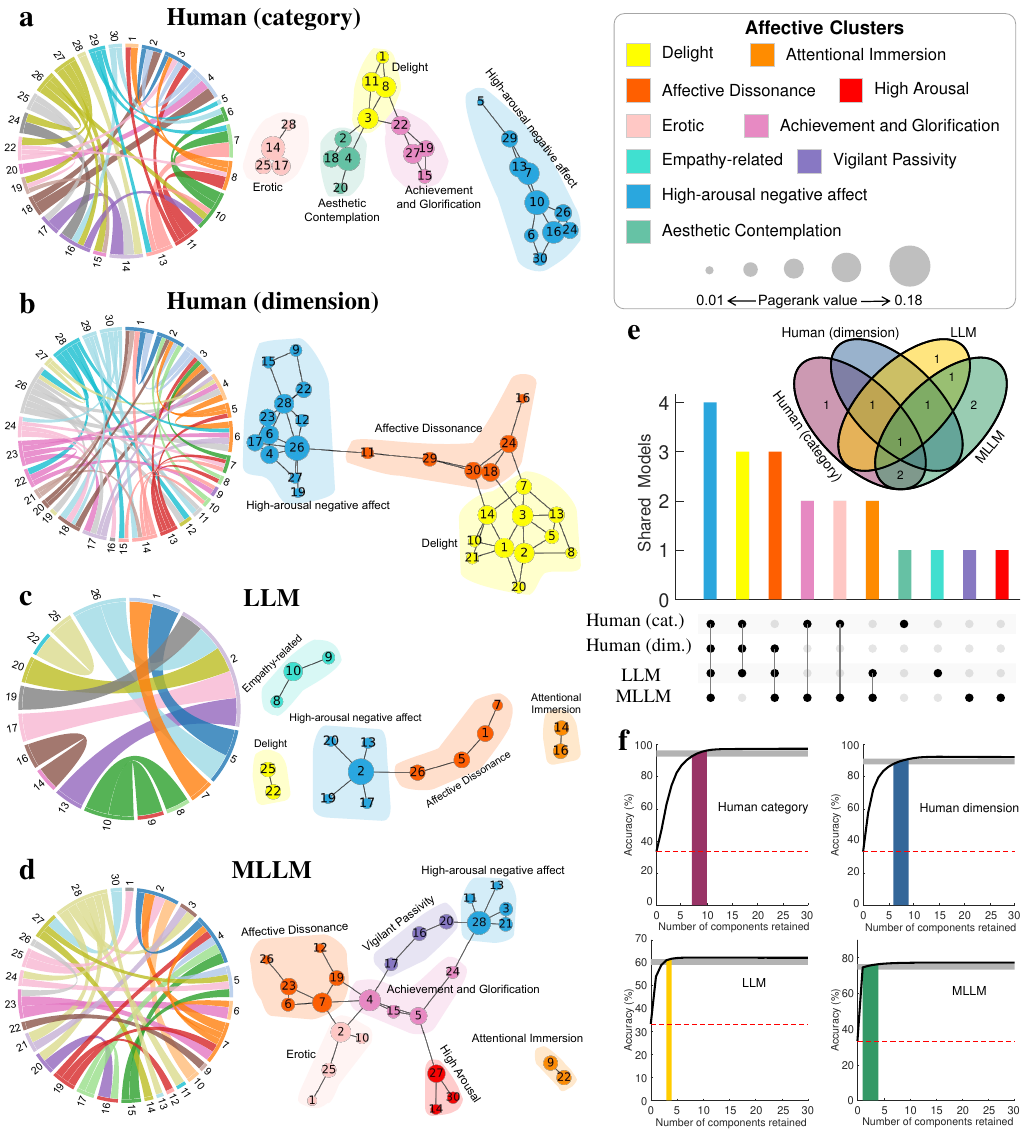}
		\caption{\textbf{Convergent and divergent structures of affective spaces across humans and models.} \textbf{a-d}, Affective graphs (left) and community structure (right) for each of the four embedding spaces. In the graphs, nodes are affective components, and edge width is proportional to the Pearson correlation between them ($r > 0.2$). In the community plots, components are grouped into clusters using the Louvain algorithm, with node size reflecting PageRank centrality. Isolated nodes are excluded. \textbf{e}, The overlap of affective clusters across all four systems, highlighting both shared and unique high-level structures. \textbf{f}, Dimensionality reduction analysis showing the minimum number of principal components required to retain 95\%–99\% of the predictive accuracy on the behavioral task (gray shaded area). The chance-level accuracy is marked by the red dashed line.}
		\label{fig:emotion-space-analyse} 
	\end{figure}

	\subsection*{Convergent evolution and divergent signatures in human and artificial affect.}
	
	To map the high-level organizational principles, we visualized the affective spaces of all four systems as graphs of correlated components and their community structures, revealing profound convergences and divergences (Fig.~\ref{fig:emotion-space-analyse}a-d). Graph analysis immediately highlighted the impact of cognitive architecture: the MLLM and human rating-derived spaces possess a richly interconnected structure, whereas the LLM’s space was significantly more fragmented, with substantially more isolated, uncorrelated components (14 for LLM vs. 1-4 for other systems). This relative isolation in the LLM likely reflects its abstract, non-grounded nature. Furthermore, community detection revealed that the MLLM developed the most granular affective structure with seven distinct clusters, whereas the human dimensional space was the most consolidated with only three. This superior granularity in the MLLM may partly explain its superior neurally-alignment. These affective communities were annotated through systematic analysis of intra-cluster label commonality (Extended Data Table \ref{tab:clusters}) and representative stimuli, aligning well with established emotion theories \cite{barrett2009affect, chatterjee2014neuroaesthetics, nadal2011copenhagen, kringelbach2010neuroscience, lambert2015foundational, ulrich2014neural, van2009neural, colosio2017neural, georgiadis2012human, panksepp2010affective, ledoux2013emotion}.

	Despite these structural differences, we identified a \textit{High-arousal negative affect} cluster \cite{ledoux2013emotion} that was remarkably conserved across all four systems---a shared continental feature in the affective landscape (Fig.~\ref{fig:emotion-space-analyse}e). This shared core, maximally responsive to threatening stimuli like car crashes and horror scenes (Extended Data Figs.~\ref{fig: LLM-MLLM_extend1}-\ref{fig: LLM-MLLM_extend3}), likely reflects its fundamental evolutionary salience in humans \cite{panksepp2004affective}, which is mirrored in the high frequency of threat-related language in the models' training data \cite{bender2021bender}. Beyond this shared core, the systems diverged in ways that again underscored the importance of sensory grounding. Specialized clusters for complex socio-cultural emotions, such as those related to \textit{Erotic} content and \textit{Achievement}, emerged in both the MLLM and human categorical representations but were notably absent in the LLM. The MLLM's ability to grasp these nuanced, embodied concepts---which the purely symbolic LLM cannot---provides compelling final evidence for the sensory grounding hypothesis. This was also reflected in the \textit{Delight} cluster, where the MLLM fractionated into more specific, fine-grained sub-clusters. Furthermore, we identified a unique \textit{Affective Dissonance} cluster in human dimensional representations, containing both positive (\textit{amusement}, \textit{joy}) and negative (\textit{fear}) embeddings while bridging the \textit{Delight} and \textit{High-arousal negative affect} clusters. This valence-ambivalent organization, absent in human categorical representations, likely represents an intermediate state along the valence continuum. Its presence in both LLM and MLLM indicates that artificial systems employ hybrid strategies beyond strict categorical or dimensional frameworks for affective processing.

	Finally, a difference emerged not in the content but in the efficiency and richness of the representations. To probe these representations, we assessed the number of components required to explain behavior. Although all systems initially represented each stimulus with 30 components, we found these representations to be highly reducible. For any given stimulus, retaining only a small subset of its most critical components was sufficient to achieve 95–99\% of the full model's predictive accuracy in the odd-one-out task, as shown by the average number of components required per stimulus in Fig.~\ref{fig:emotion-space-analyse}f.
	Crucially, the AI models required far fewer components (LLM: 3–4; MLLM: 1–4) than the human rating-derived representations (human-category: 7–10; human-dimension: 6–9). This suggests that while (M)LLMs develop human-like affective structures, their underlying representations are currently more compressed than their human counterparts.\\

	\section*{Discussion}
	
	Here, we provide compelling evidence that MLLMs can autonomously develop rich, neurally-aligned representations of human emotion. By employing (M)LLMs as indefatigable cognitive agents to perform millions of similarity judgments, we uncovered an emergent affective space that is not only highly interpretable and structurally analogous to human emotion concepts, but also strikingly predictive of neural activity. Our central finding—that an MLLM's learned representation explains brain activity in emotion-processing networks with greater accuracy than representations derived from human self-reports—offers a potential resolution to the long-standing `behavior-neural gap' in affective science. This work provides strong support for two critical hypotheses: first, that large-scale, similarity-based behavioral paradigms can capture the brain's native affective geometry more faithfully than traditional rating scales; and second, that sensory grounding is essential for developing a truly neurally-aligned conceptual framework for emotion.

	\subsection*{A machine-behavioral paradigm: MLLMs as computational cognitive agents.}
	
	A critical aspect of our study is the conceptualization of large AI models not merely as objects of study, but as revolutionary tools for psychological and neuroscientific inquiry. While some research focuses on understanding the internal mechanisms of these models by analyzing their weights and activations \cite{wang2023better, sartzetaki2024one, conwell2024large, antonello2023scaling, conwell2025perceptual}, our work pioneers a complementary approach that can be termed `large-scale machine behavior.' In this paradigm, the model serves as an indefatigable `cognitive agent,' performing behavioral tasks at a scale previously unimaginable for human participants \cite{binz2025foundation}. The innovation lies not in dissecting the model's internal parameters, but in leveraging its emergent behavioral judgments to reconstruct a high-fidelity map of a complex affective space. The resulting affective representational geometry is then validated against the gold standard of human brain activity. This positions our work as a methodological advance, demonstrating how to use the \textit{outputs} of AI to generate and test hypotheses about the structure of human emotion, thus opening a frontier for theory-driven research in the affective sciences.
	
	\subsection*{Bridging the behavior-neural gap.}
	
	A persistent challenge in affective neuroscience is the `behavior-neural gap'—the discrepancy between behavioral models of emotion derived from self-reports and the actual neural representations of affective states \cite{horikawa2020neural,koide2020distinct,du2023topographic,lettieri2024dissecting,saarimaki2025cerebral}. Our findings suggest this gap may not be an intractable problem of subjective-objective mapping, but rather a methodological artifact of the tools we use. Traditional rating scales, by forcing subjective experience into predefined and often limited categorical or dimensional labels \cite{cowen2017self}, may distort the underlying high-dimensional and blended geometry of emotion. Many behavioral studies of LLMs adopt this very paradigm, administering standardized psychological scales to compare human and model rating vectors \cite{antonello2024generative, xu2025large, bartnik2025representation, asanuma2025correspondence}. While practical, this approach imposes strong human priors, constraining LLM responses within predetermined frameworks and potentially obscuring the models' true representational capabilities. 
	
	The machine-behavioral paradigm we introduced bypasses these constraints. The process of inferring a quantitative geometry from millions of simple, qualitative similarity judgments is central to our approach's success. Each triplet odd-one-out decision, while simple, imposes a local constraint on the relative distances between three points in the affective space. By aggregating millions of such constraints, we allowed the representational structure to emerge organically, enabling the SPoSE algorithm to robustly reconstruct the global, multi-dimensional structure without the experimenter-imposed biases inherent in rating scales \cite{hebart2020revealing, mahner2025dimensions}. The resulting MLLM representation showed superior alignment with neural data, suggesting it converged on a geometry more consonant with the brain's own organizational principles. This provides the strong empirical support for the hypothesis that the `behavior-neural gap' can be bridged with more sensitive, high-throughput behavioral methods, establishing our approach as a powerful framework for exploring the computational underpinnings of affective cognition.

	\subsection*{Sensory grounding as a prerequisite for neurally-aligned affective representation.}
	
	A central debate in cognitive science concerns the necessity of sensory grounding for forming abstract concepts \cite{barsalou2008grounded,barsalou2010grounded,lecun2023large,chemero2023llms,dove2024symbol,xu2025large}. Our study offers critical evidence by testing the `neural alignment hypothesis for sensory grounding'. We found that while a language-only LLM developed a coherent affective space, a vision-language-trained MLLM achieved significantly superior alignment with human brain activity. This divergence suggests that while language alone can support the construction of a conceptual blueprint of emotion \cite{xu2025large, li2024language, schlegel2025large}, direct sensory experience is critical for developing a representational geometry that is truly aligned with the brain's affective architecture. This finding resonates with studies comparing blind and sighted individuals' knowledge of sensory-related concepts \cite{wang2020two,kim2021shared}. 
	Wang et al. reveals two forms of knowledge representation in the brain: a sensory-derived one in perceptual regions, present only in the sighted, and a language-derived, sensory-independent one in the dorsal anterior temporal lobe, present in both groups \cite{wang2020two}. Our studies map onto this framework: the LLM successfully learns the language-derived representation, explaining its considerable competence. The MLLM, however, additionally develops a functional equivalent of the sensory-derived pathway. Its superior neural alignment stems from its ability to model the complete, dual-representation architecture found in sighted individuals, where these two pathways are functionally connected. Furthermore, language effectively transmits deep, causal-explanatory knowledge, not just superficial facts. Kim et al. showed that while blind individuals are less consistent on arbitrary associative facts (e.g., "bananas are yellow"), they share a rich, "intuitive theory" of color with sighted people, acquired through language \cite{kim2021shared}. The success of LLM lies in extracting this profound structure of emotion embedded in language, masterfully reverse-engineering a conceptual model from the statistical shadow sensory experience leaves in language. Nevertheless, the MLLM's advantage demonstrates that this linguistically derived knowledge is incomplete for achieving full brain alignment. Direct sensory grounding provides a granular, object-centric representational layer necessary to mirror the complete neural architecture of human affective processing \cite{niedenthal2007embodying,barrett2008embodiment,winkielman2018dynamic}.
	
	\subsection*{A convergent hybrid coding scheme for emotion.}
	
	For decades, affective science has been dominated by the debate between discrete category theories  \cite{ekman1992argument, prinz2004emotions, panksepp2004affective, ledoux2012rethinking} and continuous dimension theories \cite{russell1980circumplex, barrett2006emotions, russell2003core, lindquist2012brain, barrett2017emotions}. Our findings offer a data-driven, computational reconciliation. The learned affective spaces, particularly in the MLLM, were neither purely categorical nor purely dimensional. Instead, they revealed a hybrid, blended coding scheme. The representations were primarily organized along a categorical backbone, spontaneously forming clusters corresponding to human emotion labels. Yet, these categories were not discrete islands; they were connected by smooth, meaningful gradients and individual components often encoded a mix of categorical and dimensional features (e.g., \textit{horror} and \textit{low-upswing}). This emergent architecture provides compelling support for the Semantic Space Theory of emotion \cite{cowen2021semantic, keltner2023semantic}, which posits that human emotion concepts are simultaneously categorical and graded. The convergence of this organizational principle in both biological intelligence and AI suggests it may be a highly efficient and fundamental solution for representing the complex landscape of affective experience.

	\subsection*{Broad applications of this study.}
	
	Our findings offer several immediate applications and broader implications. As AI models demonstrate increasingly human-like capabilities in conceptual \cite{hebart2020revealing, mahner2025dimensions, du2025human, xu2025large, cui2025large} and emotional cognition \cite{abdel2024occipital, rubin2025comparing, lai2025llms, huang2024apathetic, asanuma2025correspondence, ishikawa2025ai, zhao2024emergence, chen2024emotionqueen, sabour2024emobench, huang2023emotionally}, understanding their alignment with human cognition becomes crucial. Our work contributes a powerful paradigm for this endeavor. Early approaches often addressed alignment by extracting activations from specific model layers to predict neural activity \cite{wang2023better, sartzetaki2024one, conwell2024large, ren2024large, antonello2023scaling, conwell2025perceptual}. However, this paradigm faces challenges with contemporary large-scale proprietary models where internal activations are inaccessible, and even with open models, selecting the correct layer is non-trivial. By deriving representations from model behavior rather than internal states, our approach offers a more generalizable and principled way to systematically compare AI and human cognition.

	First, the interpretable and neurally-aligned affective components we uncovered provide a blueprint for building AI systems with more genuine emotional intelligence. These representations could be used to create empathetic human-agent interfaces and social robots capable of more nuanced interactions \cite{spezialetti2020emotion, stock2022survey}. Critically, they also open avenues in computational psychiatry. For example, they could serve as a more objective and fine-grained feature set for mental health support tools that better understand user affect, potentially aiding in the early detection of depression from behavioral cues \cite{ben2025assessing, inzlicht2024praise}. Furthermore, for conditions like autism spectrum disorder, which often involve challenges in emotion recognition, these neurally-grounded components could form the basis of novel training paradigms designed to align an individual's affective judgments with neurotypical patterns.  Besides, our two-stage attribution and editing framework serves as a generalizable tool for validating and interpreting affective representations in any AI system. It also provides a novel methodology for creating controlled, diverse affective stimuli by systematically editing neutral videos to elicit targeted emotions, addressing a common bottleneck in affective science research. Finally, the large-scale machine behavior dataset collected in this study provides a valuable benchmark for evaluating and fine-tuning the emotional alignment of future AI models, accelerating the development of systems that can safely and effectively navigate the human social-emotional world.

	\subsection*{Limitations and future directions.}	
	
	A potential limitation of machine-behavioral studies is the sensitivity of model responses to the specific structure and wording of the prompts. This raises the concern that the observed judgments might reflect artifacts of the prompt design rather than the models' intrinsic representational geometry. We explicitly addressed this issue during our data collection protocol. Initial tests revealed a strong positional bias where the MLLM preferentially selected the last item in a triplet with naive prompts. To mitigate this, we developed a text-video interleaved prompt that, as demonstrated in our supplementary analyses (Supplementary Fig. \ref{fig: odd-one-out-collection}), successfully eliminated this bias and produced an approximate uniform choice distribution (Supplementary Fig. \ref{fig: data-statistics}d). While prompt engineering remains a critical variable in AI-based cognitive science, our methodological validation ensures that the large-scale behavioral data used in this study faithfully reflects the models' emergent affective structures, thereby strengthening the reliability of our conclusions.
	
	This study has several other limitations that open promising avenues for future research. Our large-scale behavioral data collection was focused on two primary models, which, while powerful, may not represent the full diversity of AI architectures. Although our smaller-scale analysis of 22 models provides evidence for the generalizability of our findings (Extended Data Fig. \ref{fig:scale_plot}), future work should extend the large-scale analysis to a wider range of models. Furthermore, our human benchmark was simulated from prior rating data; a valuable next step would be to collect large-scale triplet judgments directly from human participants to enable a direct comparison of AI- and human-derived behavioral geometries. Finally, both the video stimuli, primarily sourced from Western cultures \cite{cowen2017self}, and the training data of the AI models may contain cultural biases. The learned affective geometry might therefore reflect a culturally-specific representation of emotion \cite{cowen2020music,cowen2024emotion}. Future research should validate our findings using cross-cultural emotional datasets to explore the universality of the emergent affective space.
	
	Looking forward, our findings suggest several exciting research directions. 
	Our generative editing framework provides a powerful tool for probing functional mechanisms. A future study could use this framework to create minimally-paired video stimuli where only a single affective component is altered. Presenting these stimuli to participants in a real-time fMRI experiment would allow for a direct test of whether manipulating a specific MLLM-derived component selectively modulates activity in the brain networks we identified. Furthermore, as noted above, the potential cultural bias in the stimuli and models necessitates research on the universality of our findings. A crucial next step would be to apply our machine-behavioral paradigm to a large-scale, cross-cultural database of emotional videos to determine which aspects of the affective geometry are culturally invariant and which are culturally specific. These future steps will be critical for building a truly comprehensive and universal model of human emotion.\\

	\section*{Methods}
	
	\noindent \textbf{Stimuli and associated data.} 
	The stimuli were 2,180 unique, emotionally evocative short video clips (mean duration $\approx$ 5 s) obtained from the dataset compiled by Cowen et al. \cite{cowen2017self} (totally 2,196 video clips, where 16 duplicate clips were removed, resulting in 2,180 unique videos). These videos were originally sourced to represent a broad spectrum of 34 distinct emotion categories and depict a diverse range of psychologically significant content, from major life events and social interactions to aesthetic experiences and survival-related threats \cite{shiota2011feeling, shiota2013beyond, fredrickson1998good, rozin2003high}. Our analyses leveraged four data modalities associated with these stimuli: (1) human behavioral ratings of the videos across 34 emotion categories and 14 affective dimensions \cite{cowen2017self}; (2) detailed textual descriptions of video content (20 unique captions per video) \cite{horikawa2024mind}; (3) human brain responses (fMRI) recorded while participants viewed the videos \cite{horikawa2020neural}; and (4) the video clips themselves.\\

	\noindent \textbf{Human behavioral ratings.} 
	Human behavioral ratings for the video stimuli were obtained from Cowen et al. \cite{cowen2017self}, where they were originally collected via Amazon Mechanical Turk (AMT). These ratings consisted of two types. First, for 34 discrete emotion categories, participants rated the extent to which each video elicited each emotion (0–100 scale); these were subsequently processed to reflect the proportion of raters who reported feeling the emotion. Second, a separate cohort of participants rated each video on 14 continuous affective dimensions using a 9-point Likert scale. For both rating types, final scores for each video were generated by averaging across participants (9–17 raters per video).\\

	\noindent \textbf{Video content descriptions.} 
	Textual descriptions of the video content were obtained from Horikawa \cite{horikawa2024mind}. These descriptions were collected on AMT following the protocol of the Microsoft COCO Captions dataset \cite{chen2015microsoft}. This process yielded 20 unique, detailed captions for each video stimulus, which served as the linguistic input for the LLMs in our study.\\
	
	\noindent \textbf{Triplet odd-one-out task.} 
	To probe the models' intrinsic representational structure, we used a triplet odd-one-out task. In each trial, a model was presented with three stimuli (either videos for MLLMs or text descriptions for LLMs) and was asked to identify which of the three was most different from the other two in terms of emotional content. Crucially, no constraints or specific criteria (e.g., valence, arousal, or specific emotion categories) were provided, allowing the models to use any features they deemed relevant. This unconstrained approach minimizes the influence of experimenter priors on the emergent representations.\\
	
	\noindent \textbf{Trial sampling strategy.}
	The total number of possible triplets from 2,180 stimuli is computationally intractable ($\binom{2180}{3} \approx 1.72 \times 10^9$). Therefore, we employed a sparse yet comprehensive sampling strategy to generate the trial set. First, all possible unique pairs of stimuli were generated ($\binom{2180}{2} = 2,375,110$ dyads). Then, for each dyad, we created three unique triplets by pairing it with three different, randomly selected stimuli from the remaining pool. This procedure resulted in a final set of 7,125,330 triplets (0.41\% coverage) that was used for the primary behavioral data collection for all systems.\\

	\noindent \textbf{Simulating human behavioral responses.} 
	To create human benchmark datasets for the triplet odd-one-out task, we simulated judgments based on the behavioral ratings from Cowen et al. \cite{cowen2017self}. This simulation was performed for each of the 7.1 million triplets. For any given triplet of stimuli (A, B, C), we calculated the pairwise cosine similarities between their vector representations (A-B, A-C, B-C). The pair with the highest similarity was considered the `non-odd' pair, and the third stimulus was designated the `odd-one-out'. This procedure was performed independently using the 34-dimensional emotion category vectors and the 14-dimensional affective dimension vectors, resulting in two distinct human behavioral datasets: `Human-category' and `Human-dimension'.\\

	\noindent \textbf{Collecting behavioral responses from MLLMs.} 
	We collected behavioral data from several MLLMs, with a primary focus on Qwen2-VL 7B \cite{wang2024qwen2} for the main analysis. For this model, we collected responses for all 7.1 million triplets. To mitigate the positional bias observed in initial tests (see Supplementary Fig. \ref{fig: odd-one-out-collection}), we used a text-video-text interleaved prompt:
	\begin{quote}
		\small
		\underline{Text input 1}: "\emph{Now we need to perform a role-playing task. Your role is an expert in analyzing the emotions conveyed in videos. Firstly, please describe the content and your evoked emotional response of each of these videos individually.}" \\
		\underline{Video input}: \{[Video\_1], [Video\_2], [Video\_3]\} \\
		\underline{Text input 2}: "\emph{Then, tell me Which clip evokes an emotional response that is noticeably different from the other two? (The answer format is "Video+ID is noticeably different from the other two") Explain the reason for this difference. /n Precautions: 1. You should focus your judgement on the emotional tone, never be influenced by the video index or location when making judgments. 2. Assume you are an emotional judgment expert with the ability to feel specific emotions for each video. Do not provide ambiguous answers. Never be influenced by the video index or location when making judgments.  3. You are not given additional constraints as to the strategy you should use.}"
	\end{quote}
	During inference, videos were resized to $256 \times 256$ pixels, and frames were sampled at a variable rate depending on video duration (see Eq. \ref{fps_eq}). Model choices were extracted from the generated text using regular expressions matching the key phrase \emph{"noticeably different"}.
	\begin{equation}
		\qquad\qquad\qquad\qquad\qquad\qquad\qquad\qquad
		\text{fps} = 
		\begin{cases} 
			2.0 & \text{if } d \leq 5 \\
			1.0 & \text{if } 5 <d \leq 15 \\
			0.5 & \text{if } 15 < d \leq 20 \\
			0.2 & \text{otherwise}
		\end{cases}
		\label{fps_eq}
	\end{equation}
	
	For the comparative analysis of model scaling (see Fig. \ref{fig:scale_plot}), we collected data from a wider range of MLLMs (Qwen2-VL 2B, 72B; Llama3.2-vision 11B; ChatGPT4o-mini) on a smaller, fully-sampled validation set of 214,500 triplets derived from 66 prototypical stimuli. These 66 prototypical stimuli were selected as the two highest-rated videos per emotion category, with duplicates consolidated.  The triplet sampling protocol comprised two stages: (1) random selection of video pairs from the 66 prototypical stimuli ($\binom{66}{2}$ possible combinations), followed by (2) augmentation with 100 randomly selected distinct videos from the complete 2,180 stimuli. This systematic approach generated 214,500 total trials ($\binom{66}{2} \times 100$), thereby ensuring comprehensive sampling coverage that includes all possible triplet combinations among the 66 core stimuli.
	Note that these 214,500 triplet judgments are not included in the training data of embedding learning stage.
	
	To estimate the noise ceiling (the upper bound of predictive performance), we measured inter-trial reliability by collecting 20 responses for each of 2,000 randomly selected triplets and calculating the mean choice consistency.\\

	\noindent \textbf{Collecting behavioral responses from LLMs.} 
	Behavioral data from LLMs were collected with a primary focus on Llama3.1 (70B) \cite{touvron2023llama}, for which we evaluated 5.5 million triplets (a sample size shown to be sufficient for stable embeddings, see Supplementary Fig. \ref{fig: data-statistics}e). The prompt was analogous to the one used for MLLMs, with video inputs replaced by their corresponding textual descriptions, which were randomly sampled from the 20 available captions for each trial:
	\begin{quote}
		\small
		\underline{Text input 1}: "\emph{Now we need to perform a role-playing task. Your role is an expert in analyzing the emotions conveyed in videos, each video is replaced by its corresponding textual description. /n Firstly, please describe your evoked emotional response of each of these videos individually. Then, tell me Which video evokes an emotional response that is noticeably different from the other two? Never be influenced by the video index when making judgments. (The answer format is "Video+ID is noticeably different from the other two.")  Explain the reason for this difference. /n Precautions: 1. You should focus your judgement on the emotional tone, never be influenced by the video index when making judgments. /n 2. Assume you are an emotional judgment expert with the ability to feel specific emotions for each video. Do not provide ambiguous answers. Never be influenced by the video index when making judgments.  /n 3. You are not given additional constraints as to the strategy you should use.}" \\
		\underline{Text input 2}: \{[Video\_1: caption\_1], [Video\_2: caption\_2], [Video\_3: caption\_3]\}
	\end{quote}
	For the comparative scaling analysis, we collected data from a range of other LLMs (Qwen2.5-instruct 7B \cite{team2024qwen2}, Qwen2-emotion 7B \cite{liu2024synthvlm}, Llama3.1 8B \cite{touvron2023llama}, Llama3.1-empathy 8B \cite{liu2024synthvlm}, Llama3.1 405B \cite{touvron2023llama}, Deepseek-R1 series (7B-671B) \cite{guo2025deepseek}) on the same 214,500-triplet validation set used for the MLLMs. This included models (Qwen2-emotion 7B and Llama3.1-empathy 8B) specifically fine-tuned on empathy datasets. Due to computational constraints, a subset of over 100,000 trials was collected for some models (Qwen2-VL 72B, ChatGPT4o\_mini, and Deepseek-R1 series).\\

	\noindent \textbf{Deriving behavioral responses from non-generative models.} 
	For models that do not have generative question-answering capabilities (e.g., CLIP \cite{radford2021learning}, ResNet \cite{he2016deep}, VideoMAE \cite{tong2022videomae}), we simulated their behavioral responses based on the similarity of their feature representations. For each stimulus, we first extracted a single feature vector from a pre-trained model. For image-based models (e.g., ResNet, CLIP-image), features from individual video frames were temporally averaged. For text-based models (e.g., CLIP-text), features were extracted from the video captions. For each triplet, the odd-one-out choice was determined by computing the pairwise cosine distances between the three feature vectors; the stimulus with the greatest average distance to the other two was selected as the odd-one-out.\\

	\noindent \textbf{Learning affective embeddings with SPoSE.} 
	We used the SPoSE algorithm \cite{hebart2020revealing, zheng2019revealing} to learn a low-dimensional, sparse, and non-negative representation for each stimulus from the triplet odd-one-out behavioral data. SPoSE models the probability that a given pair $(\hat i, \hat j)$ within a triplet $(i, j, k)$ is chosen as the most similar. This probability is modeled via a softmax function over the inner products of the stimuli's respective embedding vectors, $\bm{x}$:
	\begin{equation}\label{eq:triplet}
		\qquad\qquad\qquad\qquad
		\Pr\left((\hat i, \hat j)~\big\vert~(i,j,k)\right) = \frac{\exp(\bm{x}_{\hat i}^\top \bm{x}_{\hat j})} {\exp(\bm{x}_i^\top \bm{x}_j) + \exp(\bm{x}_i^\top \bm{x}_k) + \exp(\bm{x}_j^\top \bm{x}_k)}.
	\end{equation}
	The embeddings $\bm{x}_i$ for all stimuli $i=1...N$ are optimized by maximizing the log-likelihood of the observed behavioral choices across all $M$ trials in the dataset $\mathcal{B}$. To encourage sparsity, an L1 regularization term is added to the objective function:
	\begin{equation}\label{eq:spose}
		\qquad\qquad\qquad\qquad
		\mathcal{L} = \sum_{\{(i,j,k), (\hat i, \hat j)\} \in \mathcal{B}} {\log \left( \frac{\exp(\bm{x}_{\hat i}^\top \bm{x}_{\hat j})} {\exp(\bm{x}_i^\top \bm{x}_j) + \exp(\bm{x}_i^\top \bm{x}_k) + \exp(\bm{x}_j^\top \bm{x}_k)} \right)} + \lambda \sum_{i=1}^{N} \Vert \bm{x}_i \Vert_1,
	\end{equation}
	where $\lambda$ is the regularization hyperparameter. Optimization was performed iteratively. The learning rate ($\mathit{lr}$) was set to 0.001 for all models. The regularization coefficient ($\lambda$) was set to 0.0025 for MLLM and LLM, 0.004 for Human-category, and 0.005 for Human-dimension.\\

	\noindent \textbf{Reproducibility of embedding dimensions.} 
	To assess the stability of the learned embeddings, we performed 10 independent runs of the SPoSE algorithm for each system, each with a different random initialization. To quantify reproducibility, for each dimension in a given run, we identified the dimension in each of the ten runs with which it had the highest Pearson correlation. Following established methods \cite{hebart2020revealing, du2025human}, these correlation coefficients were Fisher z-transformed before being averaged across runs to yield a final reproducibility score for each dimension. The mean scores were then inverse-transformed.\\

	\noindent \textbf{Nearest-centroid classification.} 
	To quantitatively assess the categorical structure of the learned embeddings, we performed a classification analysis. We used a nearest-centroid classifier with a leave-one-out cross-validation scheme across 2,166 videos from 27 emotion categories (categories with <10 samples were excluded). In each fold, a category's centroid was computed as the mean embedding vector of all stimuli in that category, excluding the held-out stimulus. The held-out stimulus was then assigned to the category of the nearest centroid (Euclidean distance). Overall performance was reported as the mean top-3 classification accuracy. The empirical chance level was determined via 1,000 permutation tests where category labels were randomly shuffled.\\
	
	\noindent \textbf{fMRI data acquisition and preprocessing.} 
	The fMRI dataset was obtained from Horikawa et al. \cite{horikawa2020neural}. Data from five subjects were acquired on a 3T Siemens MAGNETOM Verio scanner as they viewed the 2,180 video clips. Stimuli were presented using a luminance-calibrated projector, and head motion was minimized with individualized bite-bars. The experiment consisted of 61 runs, each 7-10 minutes long, with functional scans acquired every 2 seconds (TR=2s).
	
	The fMRI data were preprocessed using fMRIPrep 1.2.1 \cite{esteban2019fmriprep}. The pipeline included susceptibility distortion correction, slice-timing correction (AFNI 3dTshift \cite{cox1996afni}), motion correction (FSL mcflirt \cite{jenkinson2002improved}), and coregistration to the subject's T1-weighted anatomical scan (FreeSurfer bbregister \cite{greve2009accurate}). After projection to the cortical surface, nuisance variables (baseline, linear trend, and six motion parameters) were regressed out. The resulting time-series data were compensated for a 4s hemodynamic delay, despiked (removing fluctuations >3 s.d.), averaged within each video block, and finally z-scored for each voxel.\\
	
	\noindent \textbf{fMRI regions of interest (ROIs).} 
	Cortical surfaces were reconstructed using FreeSurfer \cite{fischl2012freesurfer} and visualized using Pycortex \cite{gao2015pycortex}. Cortical ROIs were defined based on the Human Connectome Project's multimodal parcellation atlas \cite{glasser2016multi}, and included areas in the visual cortex (VC), temporo-parietal junction (TPJ), inferior parietal lobule (IPL), precuneus (PC), superior temporal sulcus (STS), temporal area TE (TE), temporal area TF (TF), medial temporal cortex (MTC), medial superior temporal (MST), insula, dorsolateral prefrontal cortex (DLPFC), dorsomedial prefrontal cortex (DMPFC), ventrolateral prefrontal cortex (VLPFC), ventromedial prefrontal cortex (VMPFC), anterior cingulate cortex (ACC), and orbitofrontal cortex (OFC). Subcortical ROIs, including the thalamus, hippocampus, hypothalamus, pallidum, brainstem, caudate, putamen, nucleus accumbens, amygdala, and cerebellum, were defined following the procedure of Horikawa et al. \cite{horikawa2020neural}.\\
	
	\noindent \textbf{Neuroimaging analyses: Voxel-wise encoding and searchlight RSA.} 
	We used two complementary methods to compare the learned representations with fMRI data: voxel-wise encoding and searchlight RSA.
	
	\noindent \underline{\textit{Voxel-wise encoding.}} We trained a regularized linear regression model (ridge regression) to predict the activity of each individual brain voxel as a linear combination of the 30 dimensions of a given representation (e.g., MLLM). Models were trained and evaluated using a 5-fold cross-validation scheme. Within each training fold, the optimal regularization parameter for each voxel was selected from 100 logarithmically spaced values ($10^{-3}$ to $10^{3}$) via a nested 6-fold cross-validation. Model performance was quantified as the Pearson correlation between predicted and actual fMRI responses on the held-out test data.
	
	\noindent \underline{\textit{Searchlight RSA.}} To optimize the signal-to-noise ratio for RSA, we first selected a subset of 100 stimuli that elicited the most reliable neural patterns. Reliability was determined using a leave-one-out voxel-wise encoding procedure: for each of the 2,180 stimuli, a model was trained on the other 2,179 stimuli and used to predict the response to the held-out stimulus. The 1,000 stimuli with the highest prediction accuracy were selected for the RSA. The searchlight analysis \cite{kriegeskorte2008representational} was then performed on this subset. For each voxel, a neural RSM was computed from the patterns of activity within a spherical searchlight (radius = 3 voxels). This neural RSM was then compared to each model's RSM using Spearman's rank correlation, resulting in a whole-brain map of representational similarity.\\

	\noindent \textbf{Cortical surface visualization.} 
	All cortical brain maps were visualized on flattened surface representations of each subject's native anatomy to allow for comprehensive inspection of the results. Cortical surface meshes were reconstructed from individual T1-weighted anatomical scans using FreeSurfer \cite{fischl2012freesurfer}. The functional data were then aligned to these surfaces and plotted using Pycortex \cite{gao2015pycortex}.\\

	\noindent \textbf{Measuring human-model behavioral consistency.} 
	To compare the behavioral patterns of different models to humans (Fig. \ref{fig:scale_plot}), we used the validation set of 214,500 triplets derived from 66 prototypical stimuli. For each model and for the human benchmark data, we constructed a $66 \times 66$ RSM, where each entry represented the choice probability for a given pair of stimuli. Human-model consistency was then quantified as the Pearson correlation between the model's RSM and the human RSM, adjusted for the reliability of each RSM using the correction formula from Rajalingham et al. \cite{rajalingham2018large}. Reliability was estimated using a split-half correlation on the behavioral trials. The calculation formula is expressed as follows:
	\begin{equation}
		\qquad\qquad\qquad\qquad\qquad
		\tilde{\rho}(m, h) = \frac{\rho(\text{RSM}_m, \text{RSM}_h)}{\sqrt{\rho(\text{RSM}_m^{\text{half}_1}, \text{RSM}_m^{\text{half}_2}) \cdot \rho(\text{RSM}_h^{\text{half}_1}, \text{RSM}_h^{\text{half}_2})}},
	\end{equation}
	where $\text{RSM}_m^{\text{half}_1}$ and $\text{RSM}_m^{\text{half}_2}$ were constructed from split-half behavioral data of the 66 prototypical stimuli, with analogous computation for the human RSM components.\\

	\noindent \textbf{Labeling of the learned affective components.} 
	To interpret the meaning of each of the 30 learned components, we correlated each component's vector (a 2,180-element vector of stimulus weights) with the corresponding human rating vectors for the 34 emotion categories and 14 affective dimensions from Cowen et al. \cite{cowen2017self}. A component was assigned the labels of the top three most highly correlated human-rated terms (based on absolute Pearson's $r$), subject to a cutoff rule: if the top-correlated term's coefficient was at least 0.1 greater than the second, only the top term was kept. This process was iterated for the second and third terms. Component's labels were suffixed with `-high' or `-low' to indicate the sign of the correlation (e.g., \textit{valence-high} or \textit{arousal-low}). The full list of labels is provided in Extended Data Table \ref{table:30dim_labels}.\\

	\noindent \textbf{Attribution analysis using Grad-CAM.} 
	To identify the visual features driving each affective component, we employed an attribution analysis based on Grad-CAM \cite{selvaraju2017grad}. First, to bridge the learned affective space with a vision model's feature space, we trained a simple linear model to project video features (temporally averaged from the frame-wise features) from the pre-trained CLIP visual encoder into our 30-dimensional SPoSE space. This model was optimized via mean squared error minimization. With this projection in place, for any target affective component and input video frame, we computed the gradient of the component's activation with respect to the feature maps of the CLIP encoder's final convolutional layer. These gradients were then used to produce a saliency heatmap, which was upsampled and overlaid on the original frame to visualize the specific image regions responsible for activating that component (see Extended Data Fig. \ref{fig: GradCAM_figure}).\\
	
	\noindent \textbf{Probing the functional role of components via generative emotion editing.} 
	To probe the relationship between the learned affective components and the visual content of videos, we developed a generative editing framework that operates by mapping between our affective space and a pre-trained semantic space (Extended Data Fig. \ref{fig:emotion-edit-figure}). The core of this framework is an `emotion projector' model, which learns to translate manipulations in the affective space into targeted changes in the semantic space, which then guide a video generation model.

	\noindent \underline{\textit{Emotion projector training.}}
	The emotion projector is a five-layer MLP with LayerNorm, trained to map a 30-dimensional SPoSE embedding ($\mathbf{f}_{\text{SPoSE}}$) to CLIP's 768-dimensional joint latent space. The training objective was to align the projector's output, $\mathbf{f}_{\text{proj}} = \text{MLP}_{\text{LN}}(\mathbf{f}_{\text{SPoSE}})$, with the corresponding text and vision embeddings from the original video, using a tri-modal contrastive loss:
	\begin{equation}\label{eq:clip_loss}
		\qquad\qquad\qquad\qquad
		\mathcal{L}_{\text{CLIP}} = L_{\text{BiInfoNCE}}(\mathbf{f}_{\text{proj}}, \mathbf{f}_{\text{vision}}) + L_{\text{BiInfoNCE}}(\mathbf{f}_{\text{proj}}, \mathbf{f}_{\text{language}}),
	\end{equation}
	\begin{equation}\label{equ1}
		\qquad\qquad\qquad\qquad
		\mathcal{L}_{BiInfoNCE} = -\frac{1}{B} \sum_{i = 1}^{B} \Big( log \frac{exp(s(\mathbf{\hat{z}}_{i}, \mathbf{z}_{i}) / \tau)}{\sum_{j=1}^{B} exp(s(\mathbf{\hat{z}}_{i}, \mathbf{z}_{j}) / \tau)    }     +   log \frac{exp(s(\mathbf{\hat{z}}_{i}, \mathbf{z}_{i}) / \tau)}{\sum_{k=1}^{B} exp(s(\mathbf{\hat{z}}_{k}, \mathbf{z}_{i}) / \tau)    }                  \Big),
	\end{equation}
	where $s(\cdot,\cdot)$ is the cosine similarity, $\mathbf{z}$ and $\mathbf{\hat{z}}$ are  representation from the two modalities, $B$ is the batch size, and $\tau$ is a learned temperature parameter. For training stability, MSE loss directly supervises $\mathbf{f}_{\text{proj}}$ against $\mathbf{f}_{\text{language}}$:
	\begin{equation}\label{eq:proj1}
		\qquad\qquad\qquad\qquad\qquad\qquad\qquad\qquad
		\mathcal{L}_{\text{proj1}} = \|\mathbf{f}_{\text{proj}} - \mathbf{f}_{\text{language}}\|_2^2 .
	\end{equation}
	To enable token-level editing, we supervise an up-projected representation against token features:
	\begin{equation}\label{eq:proj2}
		\qquad\qquad\qquad\qquad
		\mathcal{L}_{\text{proj2}} = \|(\mathbf{W}_{\text{up}}  \cdot \mathbf{f}_{\text{proj}} + \mathbf{b}_{\text{up}}) - \mathbf{f}_{\text{text-token}}\|_2^2, \quad \mathbf{W}_{\text{up}} \in \mathbb{R}^{768 \times (77 \times 768)} .
	\end{equation}
	The composite loss combines these objectives with adjustable hyperparameters:
	\begin{equation}\label{eq:total_loss}
		\qquad\qquad\qquad\qquad\qquad\qquad\qquad\qquad
		\mathcal{L} = \mathcal{L}_{\text{proj1}} + \lambda_1  \cdot \mathcal{L}_{\text{CLIP}} +   \lambda_2  \cdot \mathcal{L}_{\text{proj2}} .
	\end{equation}
	The model was trained for 600 epochs, with a 50-epoch warm-up period, using an Adam optimizer, a learning rate of $2 \times 10^{-4}$, and a batch size of 128. The loss weighting hyperparameters were set to $\lambda_1 = 0.5$ and $\lambda_2 = 1$.

	\noindent \underline{\textit{Video editing inference pipeline.}}
	The editing process at inference proceeded as follows. First, a target dimension in a video's original SPoSE embedding ($\mathbf{f}_{\text{SPoSE}}$) was manually modified to create an edited embedding ($\mathbf{\hat{f}}_{\text{SPoSE}}$). This edited embedding was then passed through the trained emotion projector to generate a new set of target semantic tokens. In parallel, the original video was encoded into a latent representation by a VQVAE encoder. DDIM inversion was applied to this latent representation to generate an initial noise tensor that preserved the temporal structure and content of the original video. Finally, a pre-trained Stable Diffusion model (v1.5) was conditioned on the new target semantic tokens to denoise this initial tensor, generating the edited video frame by frame in the latent space. The edited latent frames were then reconstructed into pixel space by the VQVAE decoder to produce the final output video.\\

	\noindent \textbf{Emotion graph construction and analysis.} 
	To analyze the high-level structure of the learned affective spaces, we first constructed an `emotion graph' for each system. In these graphs, the 30 learned components were treated as nodes. An edge was drawn between any two nodes if the absolute Pearson correlation of their stimulus weights exceeded a threshold of $\rho > 0.2$, with the edge weight set to the correlation value. To identify clusters of related components, we applied the standard Louvain community detection algorithm to partition the graph into non-overlapping communities \cite{blondel2008fast}. We removed the resulting isolated nodes. Finally, to identify the most central or influential components within each graph, we calculated the PageRank centrality for each node \cite{brin1998anatomy}.

	\section*{Data availability}
	
	The short videos and their human ratings can be requested from \url{https://goo.gl/forms/XErJw9sBeyuOyp5Q2}. 
	The preprocessed fMRI: \url{https://github.com/KamitaniLab/EmotionVideoNeuralRepresentationPython}.
	The video captions are available at \url{https://github.com/horikawa-t/MindCaptioning}.
	The machine-behavioral judgments can be found at \url{https://osf.io/wtp98/?view_only=0ad8badccb2a4bbcac5c887f53e96238}.

	\section*{Code availability}
	
	The codes used for data collection, embedding learning, result analysis, and visualization in this study are publicly available at \url{https://osf.io/wtp98/?view_only=0ad8badccb2a4bbcac5c887f53e96238}.

	\bibliography{sample}

	\section*{Acknowledgements} 
	This work was supported in part by the Strategic Priority Research Program of the Chinese Academy of Sciences (Grant No. XDB1010202); in part by the National Natural Science Foundation of China under Grant 62020106015 and Grant 62206284; in part by Beijing Natural Science Foundation under Grant L243016, and in part by the Beijing Nova Program under Grant 20230484460.
	We would like to thank Alan S. Cowen and Dacher Keltner for sharing the emotionally evocative video database and the human behavioral ratings on 34 emotion categories and 14 affective dimensions.  We also thank Yukiyasu Kamitani and Tomoyasu Horikawa for sharing the corresponding fMRI data and video captions. We also thank Martin N. Hebart, Jack L. Gallant, et al. for sharing the code and tools for embedding learning and visualization analysis.

	\section*{Author contributions}
	
	C.D., Y.L. and H.H. designed the research; Y.L., Z.H. and C.D. conducted the experiments; Y.L. and Z.H. collected the data; Y.L., C.D. and Z.H. wrote the paper;  all the authors analyzed the results;  all the authors approved the paper.
	
	\section*{Competing interests} 
	
	The authors declare no competing interests.
	
	\section*{Additional information}  
	Correspondence and requests for materials should be addressed to H.H.

	\section*{Extended data}
	
	\beginextended
	\extendeddatafigure	
	\extendeddatatable

	\begin{figure}[!h]
		\centering
        \includegraphics[width=17.5cm]{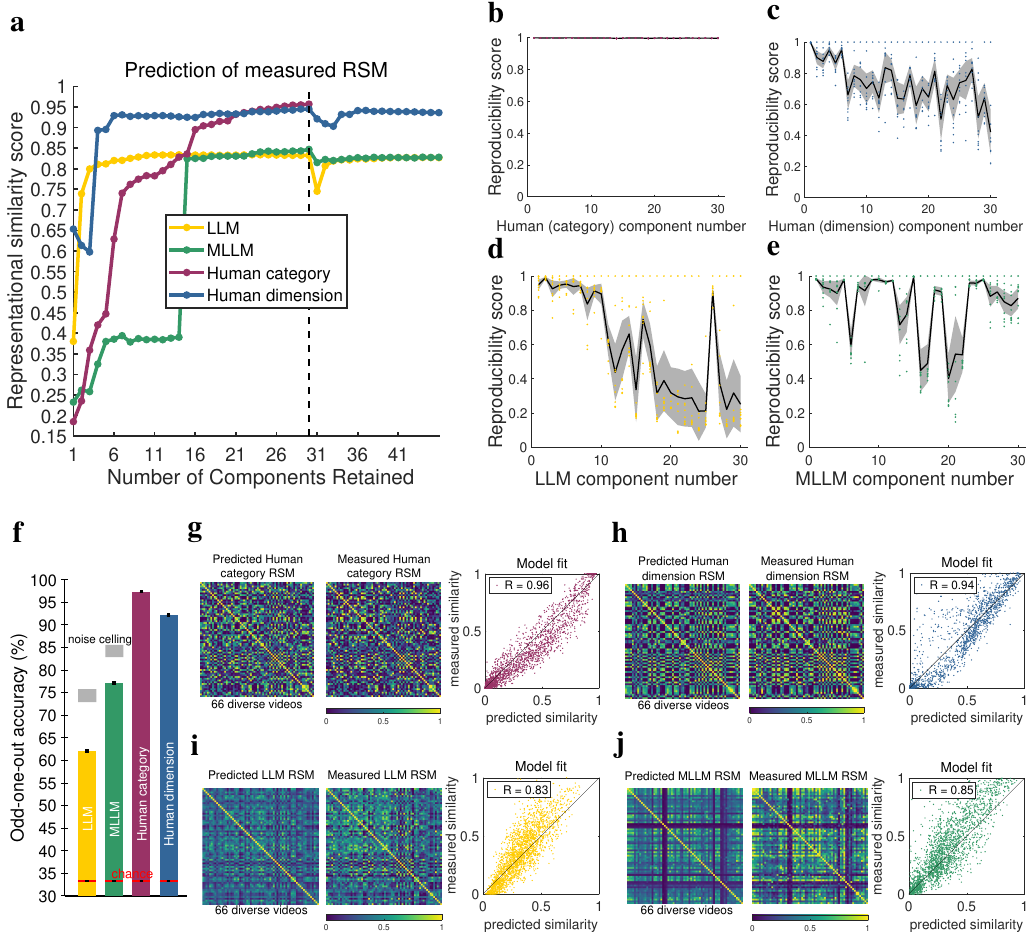}
		\caption{\textbf{Learned affective embeddings are stable and predictive.} \textbf{a}, Effect of dimensionality on the ability of embeddings to predict held-out similarity judgments, showing performance saturation at approximately 30 components. \textbf{b-e}, Reproducibility of the 30 learned components across ten independent model runs. Each point is the maximum correlation of a component from one run with all components from the other runs. Shaded areas represent 95\% confidence intervals (CIs). \textbf{f}, Prediction accuracy of the 30-dimensional embeddings on a held-out set of triplet choices. Noise ceilings (upper bound of performance) were estimated from trial-to-trial reliability (across 10,000 triplets). Error bars are 95\% CIs. \textbf{g-j}, Comparison of RSMs for 66 validation stimuli. Left: RSMs predicted by the learned embeddings. Middle: RSMs measured from empirical behavioral choices (unavailable in the SPoSE training). Right: Pearson's correlation between predicted and measured RSMs, indicating high global correspondence.}
		\label{fig:Dimensional-Validity}
	\end{figure}

	\begin{figure}[!h]
		\centering
        \includegraphics[width=17.5cm]{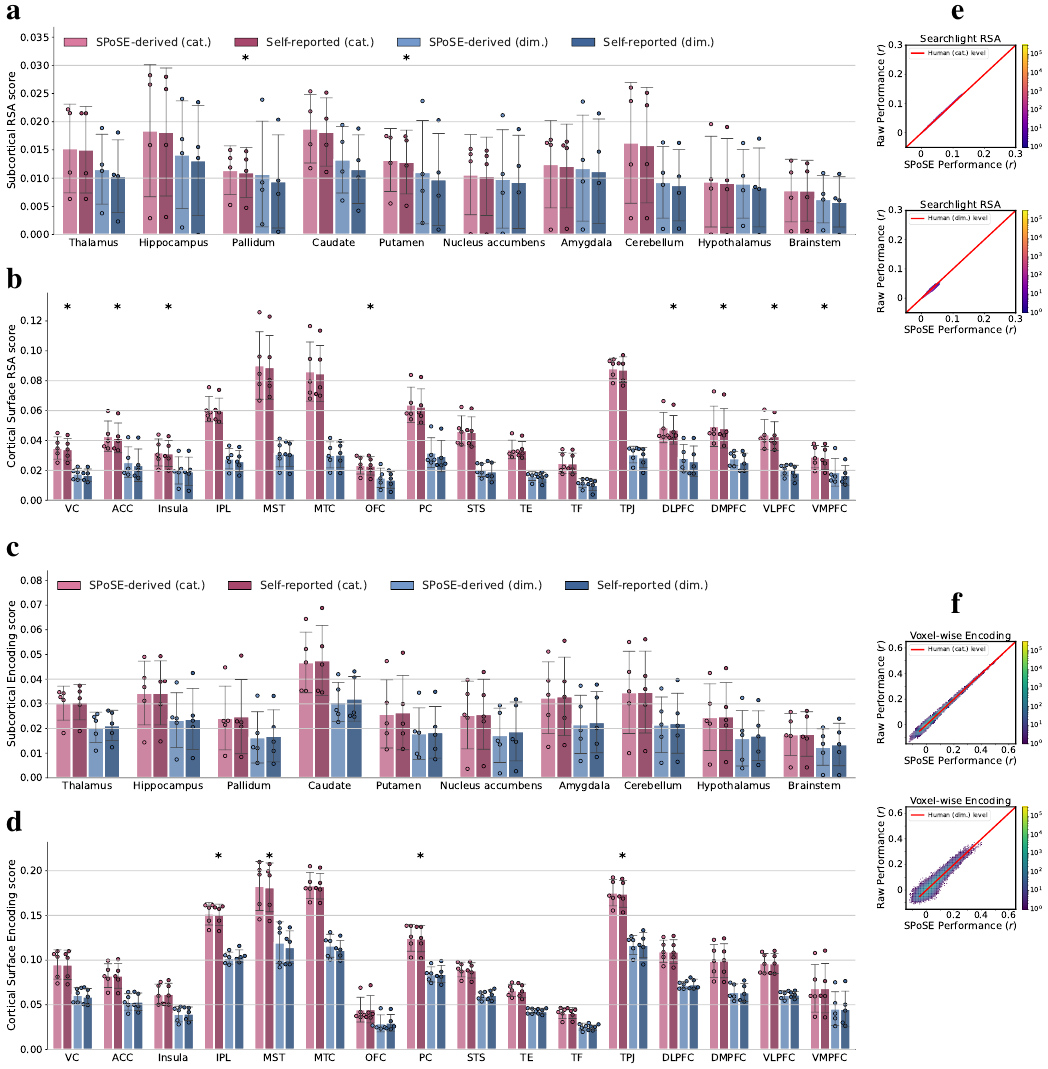}
		\caption{\textbf{SPoSE-derived human embeddings show comparable neural alignment to original self-report ratings.}  \textbf{a, b}, Comparison of neural alignment using searchlight RSA, averaged across subcortical (\textbf{a}) and cortical (\textbf{b}) ROIs. \textbf{c, d}, Comparison of neural alignment using voxel-wise encoding, averaged across subcortical (\textbf{c}) and cortical (\textbf{d}) ROIs. Performance (Pearson's $r$) is averaged across subjects (N=5), with individual subjects shown as dots and error bars representing standard deviation (s.d.). The results indicate that the performance of the SPoSE-derived human embeddings is comparable to their corresponding self-report rating vector counterparts, with few statistically significant differences observed (marked by stars, $P$ < 0.05, FDR corrected). \textbf{e, f}, 2D histogram comparison for a representative subject (Subject 1), plotting the performance of the SPoSE-derived human embeddings against the original self-report ratings. Results are shown for both searchlight RSA (\textbf{e}) and voxel-wise encoding (\textbf{f}). The voxels are distributed symmetrically around the identity line, indicating that both methods capture neural information to a similar degree across the brain. }
		\label{fig:spose_vs_raw_comparison} 
	\end{figure}

	\begin{figure}[!h]
		\centering
        \includegraphics[width=17.5cm]{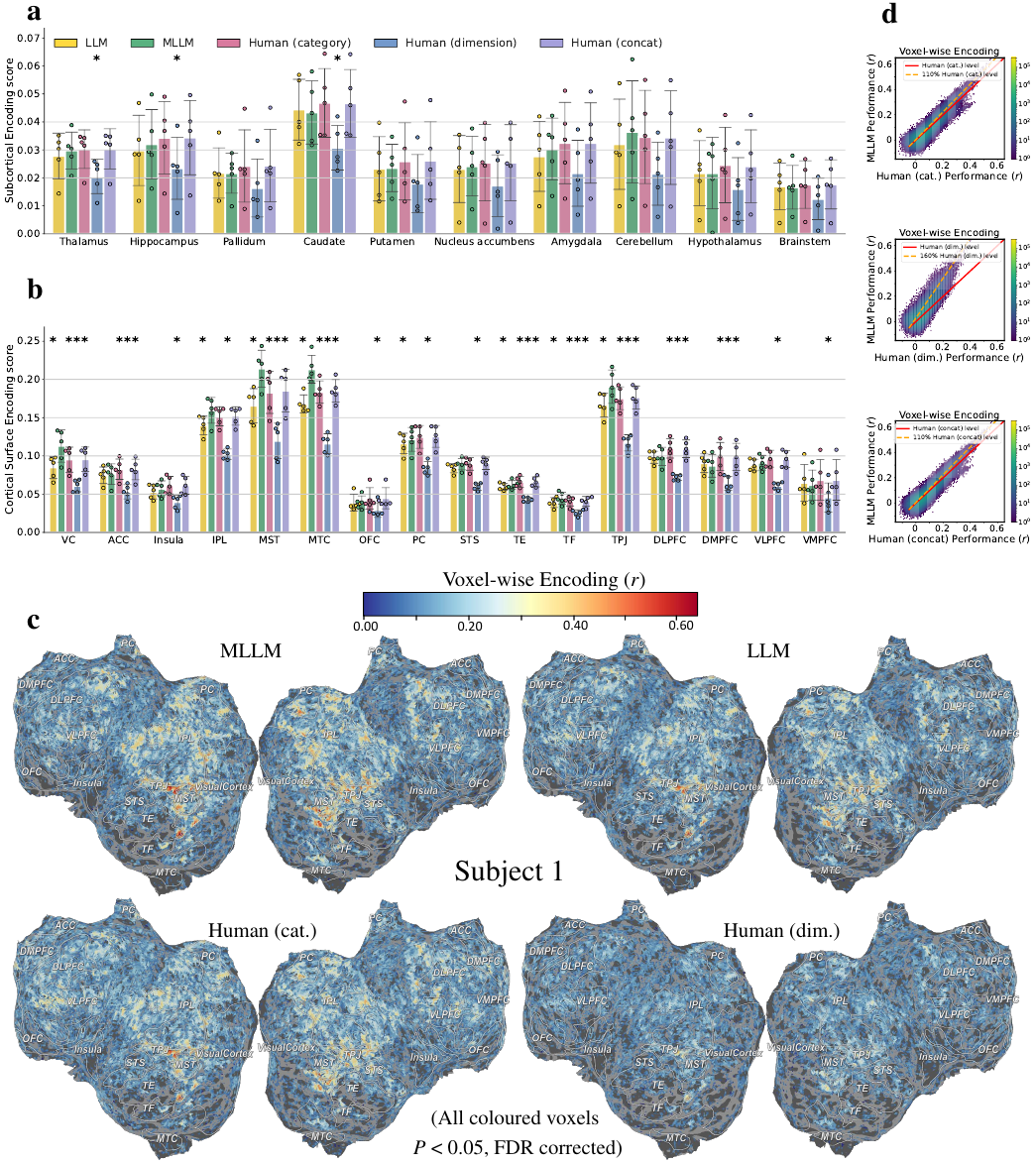}	
		\caption{\textbf{Results of voxel-wise encoding.} Prediction accuracy (Pearson's $r$) averaged for subcortical (\textbf{a}) and cortical (\textbf{b}) ROIs (averaged across subjects (N=5)). Dots represent individual subjects; error bars reflect standard deviation; all statistics are two-tailed t-tests across subjects, with FDR correction; stars indicate significant differences between MLLM and the compared model ($P$ < 0.05). \textbf{c}, Flattened cortical maps of voxel-wise encoding accuracy (Pearson's $r$) for Subject 1. All coloured voxels are predicted significantly ($P$ < 0.05, FDR-corrected, two-tailed t-tests). \textbf{d}, Voxel-wise comparison of the MLLM's performance against the human categorical, dimensional and concatenated ratings model for Subject 1. }
		\label{fig:RSA_encoding_appendix} 
	\end{figure}

	\begin{figure}[!h]
		\centering
        \includegraphics[width=17.4cm]{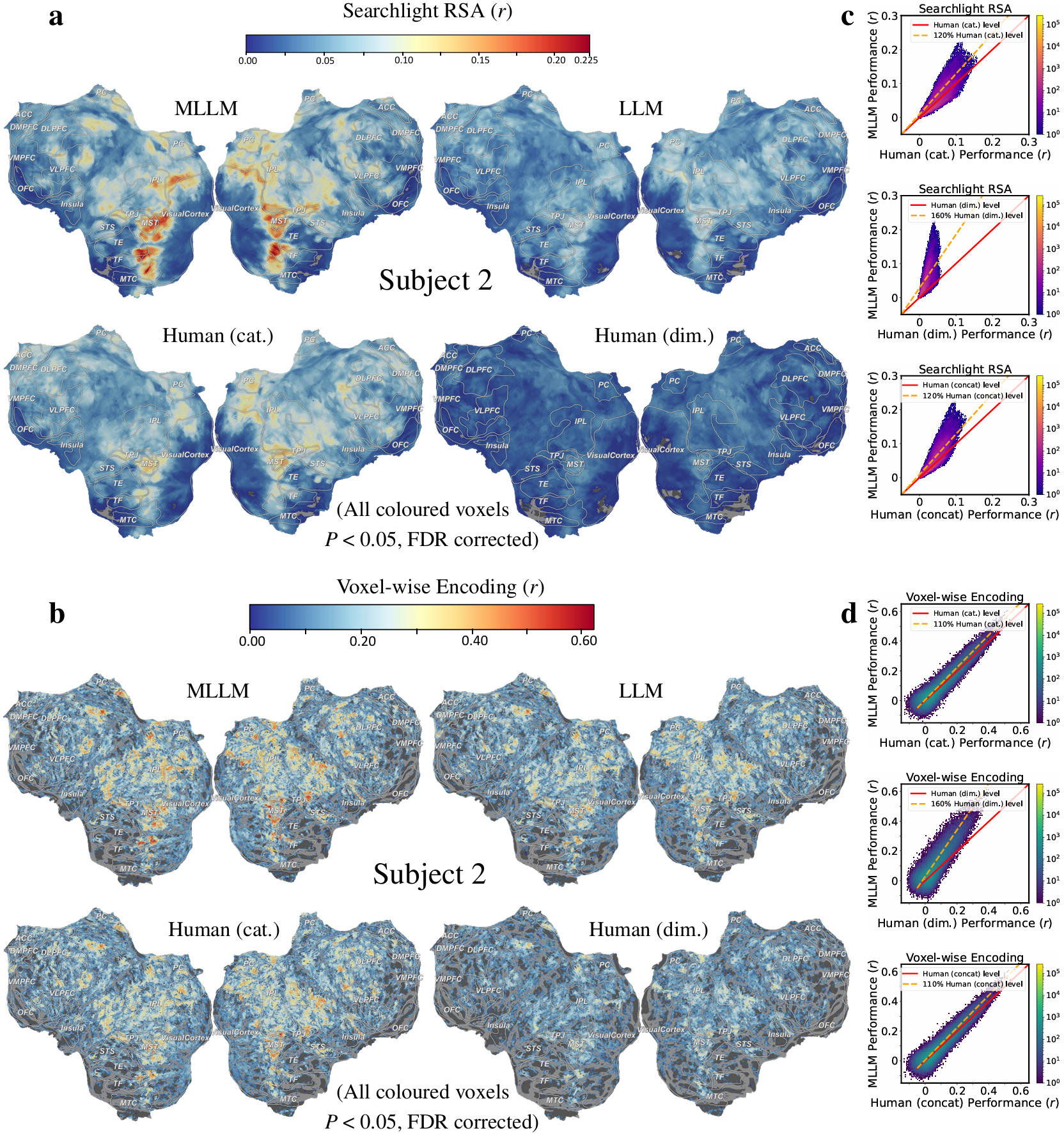}
		\caption{\textbf{Results of searchlight RSA and voxel-wise encoding for Subject 2. }\textbf{a, b} Flattened cortical maps of searchlight RSA (\textbf{a}) and voxel-wise encoding (\textbf{b}). All coloured voxels are predicted significantly ($P$ < 0.05, FDR-corrected, two-tailed t-tests). \textbf{c, d}, Voxel-wise comparison of the MLLM's performance against the human categorical, dimensional and concatenated ratings model, using searchlight RSA (\textbf{c}) and voxel-wise encoding (\textbf{d}).}
		\label{fig:sub2_cortex} 
	\end{figure}

	\newpage

	\begin{figure}[!h]
		\centering
        \includegraphics[width=17.4cm]{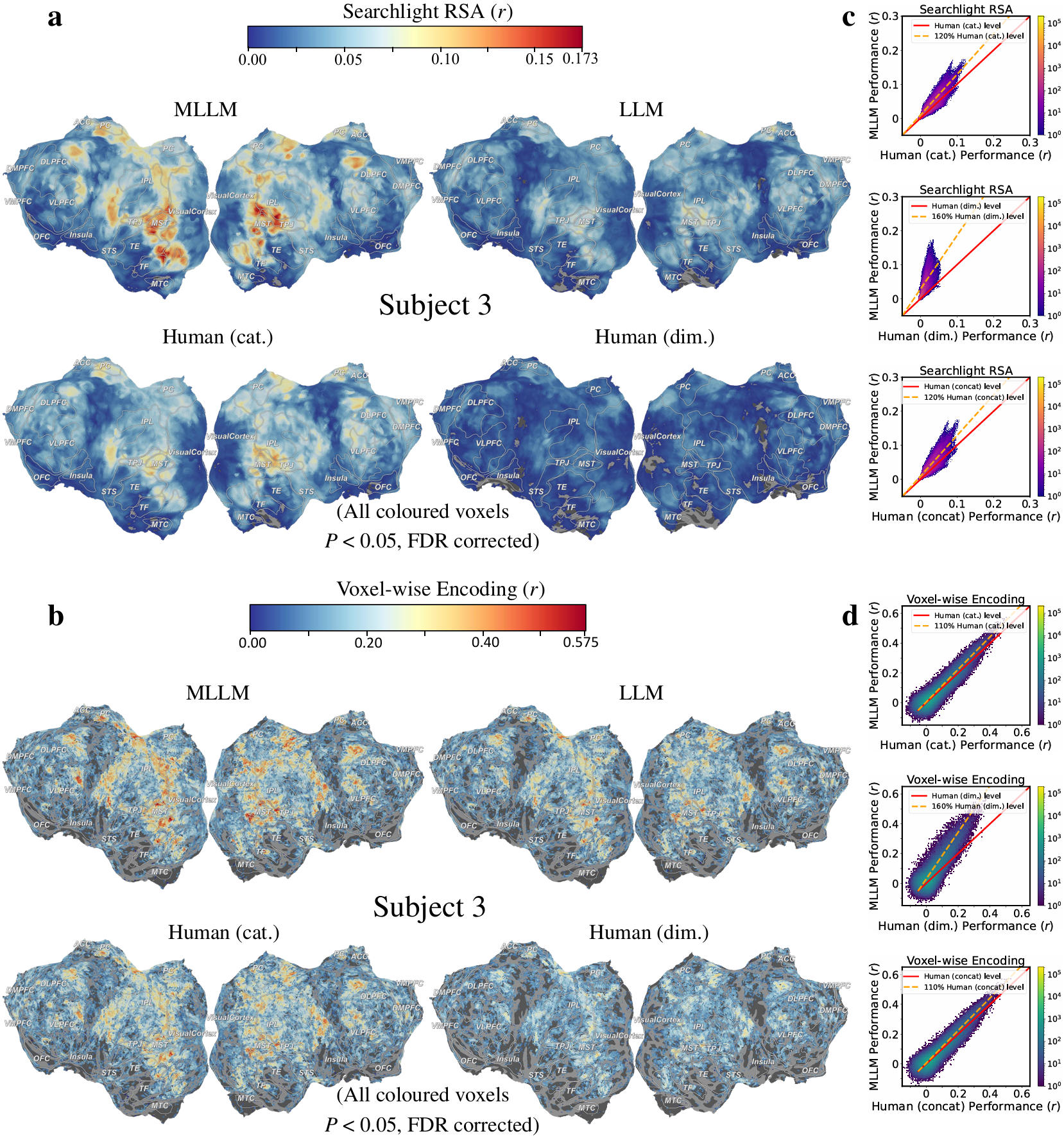}
		\caption{\textbf{Results of searchlight RSA and voxel-wise encoding for Subject 3. }\textbf{a, b} Flattened cortical maps of searchlight RSA (\textbf{a}) and voxel-wise encoding (\textbf{b}). All coloured voxels are predicted significantly ($P$ < 0.05, FDR-corrected, two-tailed t-tests). \textbf{c, d}, Voxel-wise comparison of the MLLM's performance against the human categorical, dimensional and concatenated ratings model, using searchlight RSA (\textbf{c}) and voxel-wise encoding (\textbf{d}).}
		\label{fig:sub3_cortex} 
	\end{figure}

	\newpage

	\begin{figure}[!h]
		\centering
        \includegraphics[width=17.4cm]{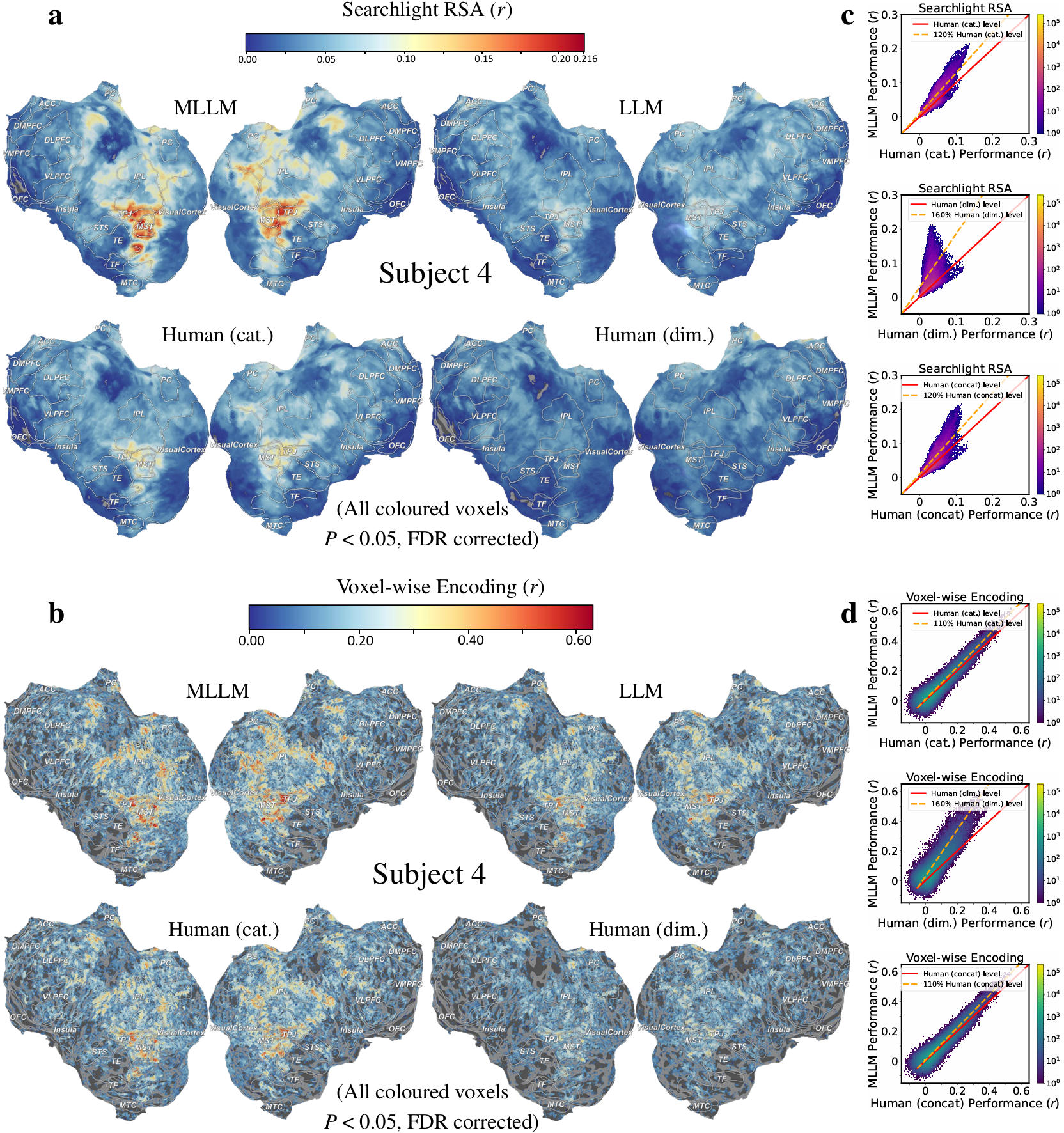}
		\caption{\textbf{Results of searchlight RSA and voxel-wise encoding for Subject 4. }\textbf{a, b} Flattened cortical maps of searchlight RSA (\textbf{a}) and voxel-wise encoding (\textbf{b}). All coloured voxels are predicted significantly ($P$ < 0.05, FDR-corrected, two-tailed t-tests). \textbf{c, d}, Voxel-wise comparison of the MLLM's performance against the human categorical, dimensional and concatenated ratings model, using searchlight RSA (\textbf{c}) and voxel-wise encoding (\textbf{d}).}
		\label{fig:sub4_cortex} 
	\end{figure}

	\newpage

	\begin{figure}[!h]
		\centering
        \includegraphics[width=17.4cm]{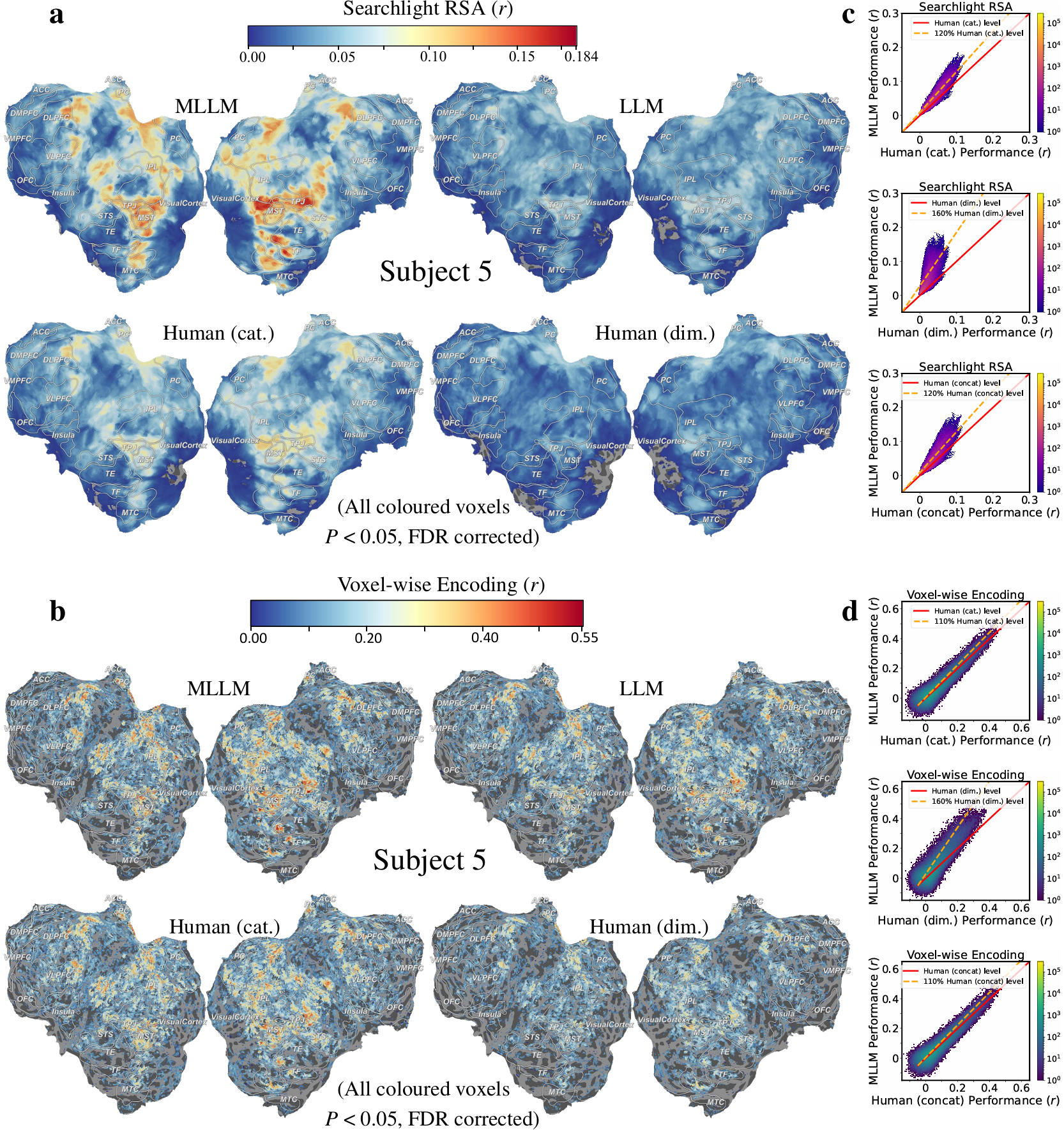}
		\caption{\textbf{Results of searchlight RSA and voxel-wise encoding for Subject 5. }\textbf{a, b} Flattened cortical maps of searchlight RSA (\textbf{a}) and voxel-wise encoding (\textbf{b}). All coloured voxels are predicted significantly ($P$ < 0.05, FDR-corrected, two-tailed t-tests). \textbf{c, d}, Voxel-wise comparison of the MLLM's performance against the human categorical, dimensional and concatenated ratings model, using searchlight RSA (\textbf{c}) and voxel-wise encoding (\textbf{d}).}
		\label{fig:sub5_cortex} 
	\end{figure}

	\newpage
	\begin{figure}[!h]
		\centering	
        \includegraphics[width=17.4cm]{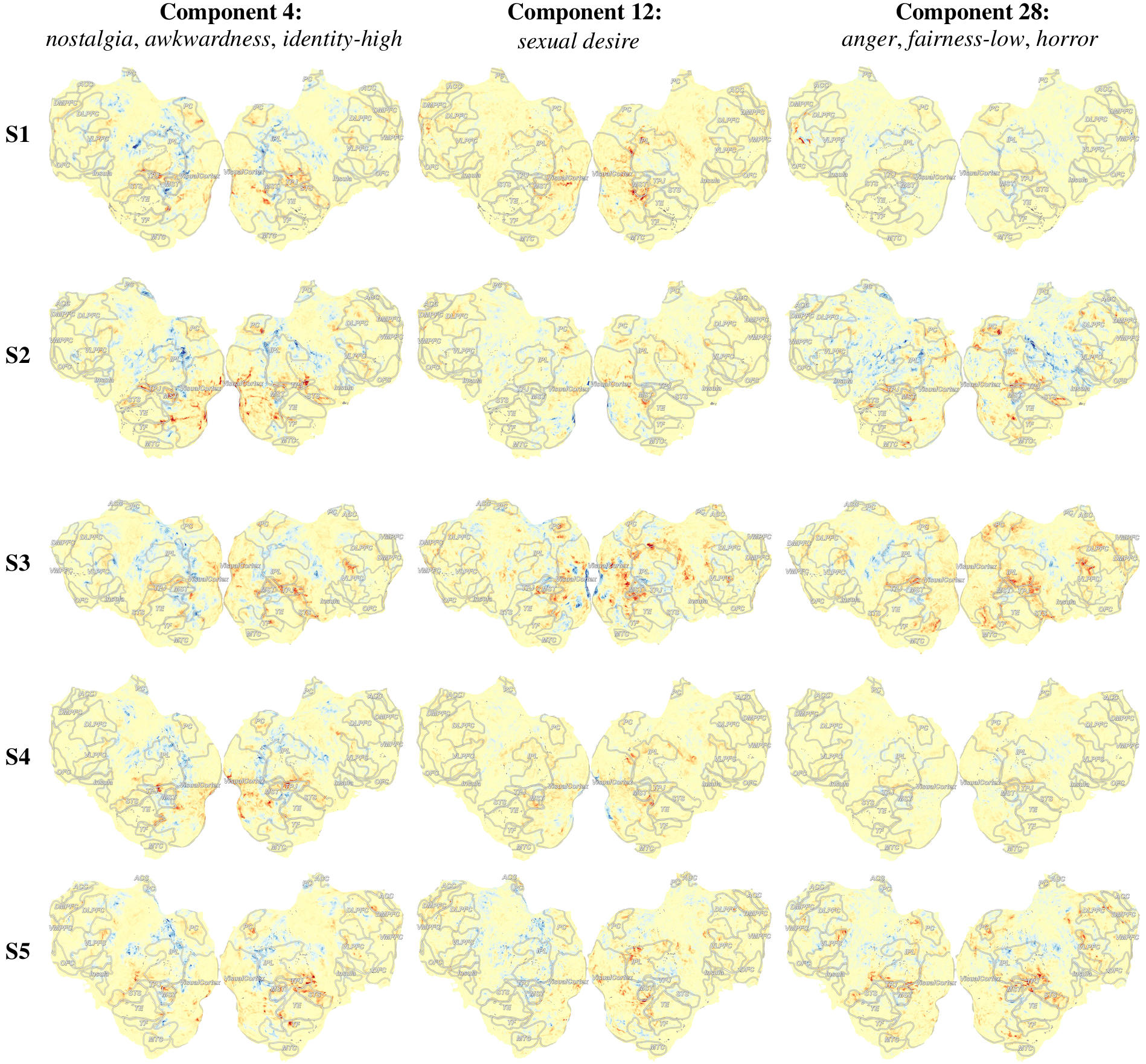}
		\caption{\textbf{Neural correlates of specific MLLM affective components.} This figure shows the brain activation patterns triggered by three distinct, interpretable affective components from the MLLM, generated from the weights of the voxel-wise encoding models. Each row displays the results for a single subject (S1–S5), and each column corresponds to a specific component (e.g., \textit{nostalgia}, \textit{sexual desire}, \textit{anger}). Red indicates positive weights, reflecting brain regions where activity is positively associated with the affective component; blue indicates negative weights, reflecting regions with a negative association. The high degree of consistency in the spatial patterns across all five subjects suggests that these learned affective components capture neurally meaningful and stable components of emotion processing.}
		
		\label{fig:dim_vis} 
		
	\end{figure}

	\newpage
	\begin{figure}[!h]
		\centering
        \includegraphics[width=16.8cm]{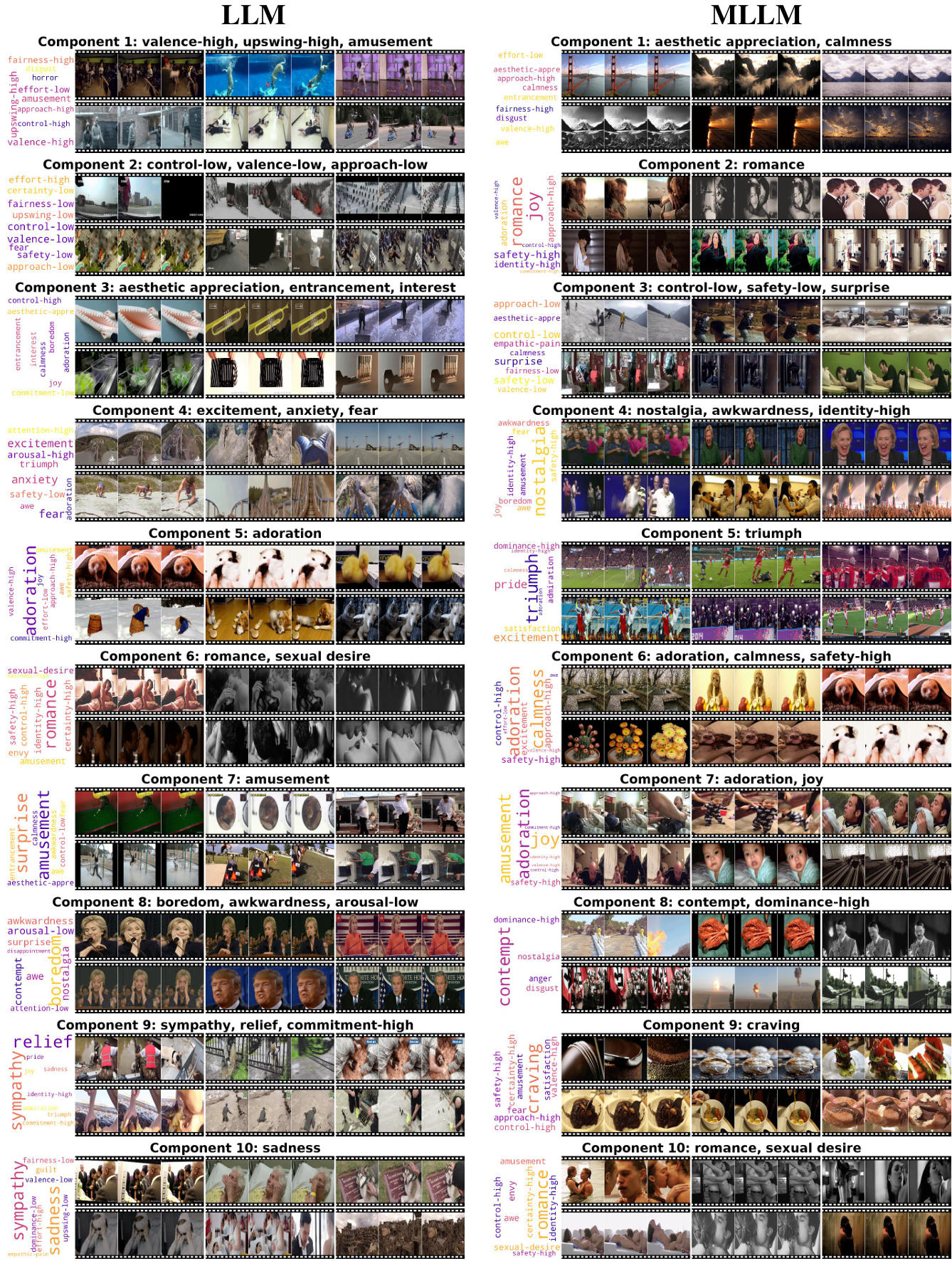}
		\caption{\textbf{Visualization of learned affective components 1–10 for the LLM and MLLM (related to Fig. \ref{fig:Dimensional-Vis}).} For each component, we display its assigned affective label(s) and the six video frames that received the highest weight. }
		
		\label{fig: LLM-MLLM_extend1}
	\end{figure}

	\begin{figure}[!h]
		\centering
        \includegraphics[width=16.8cm]{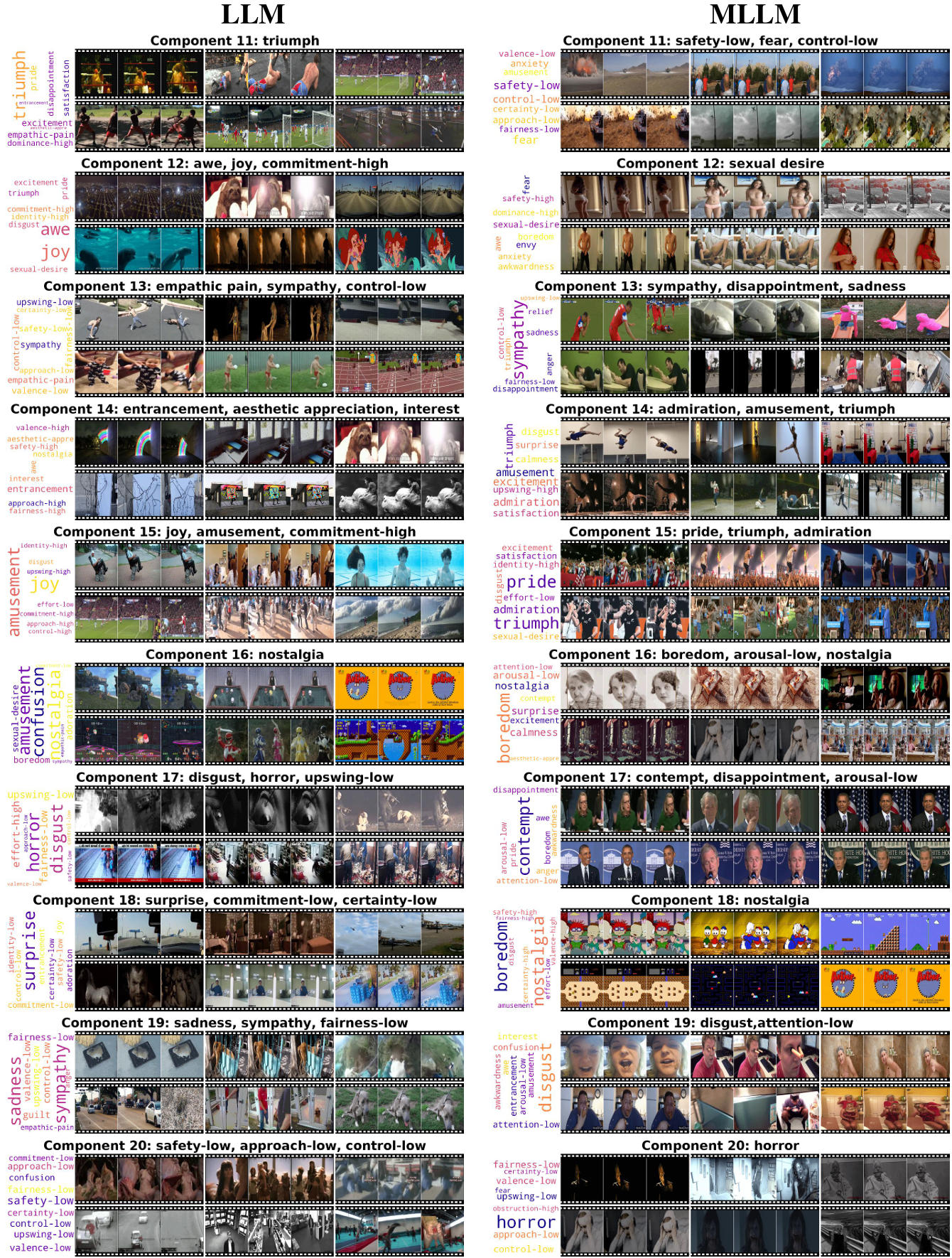}
		\caption{\textbf{Visualization of learned affective components 11–20 for the LLM and MLLM (related to Fig. \ref{fig:Dimensional-Vis}).} For each component, we display its assigned affective label(s) and the six video frames that received the highest weight. }
		\label{fig: LLM-MLLM_extend2}
	\end{figure}

	\begin{figure}[!h]
		\centering
        \includegraphics[width=16.8cm]{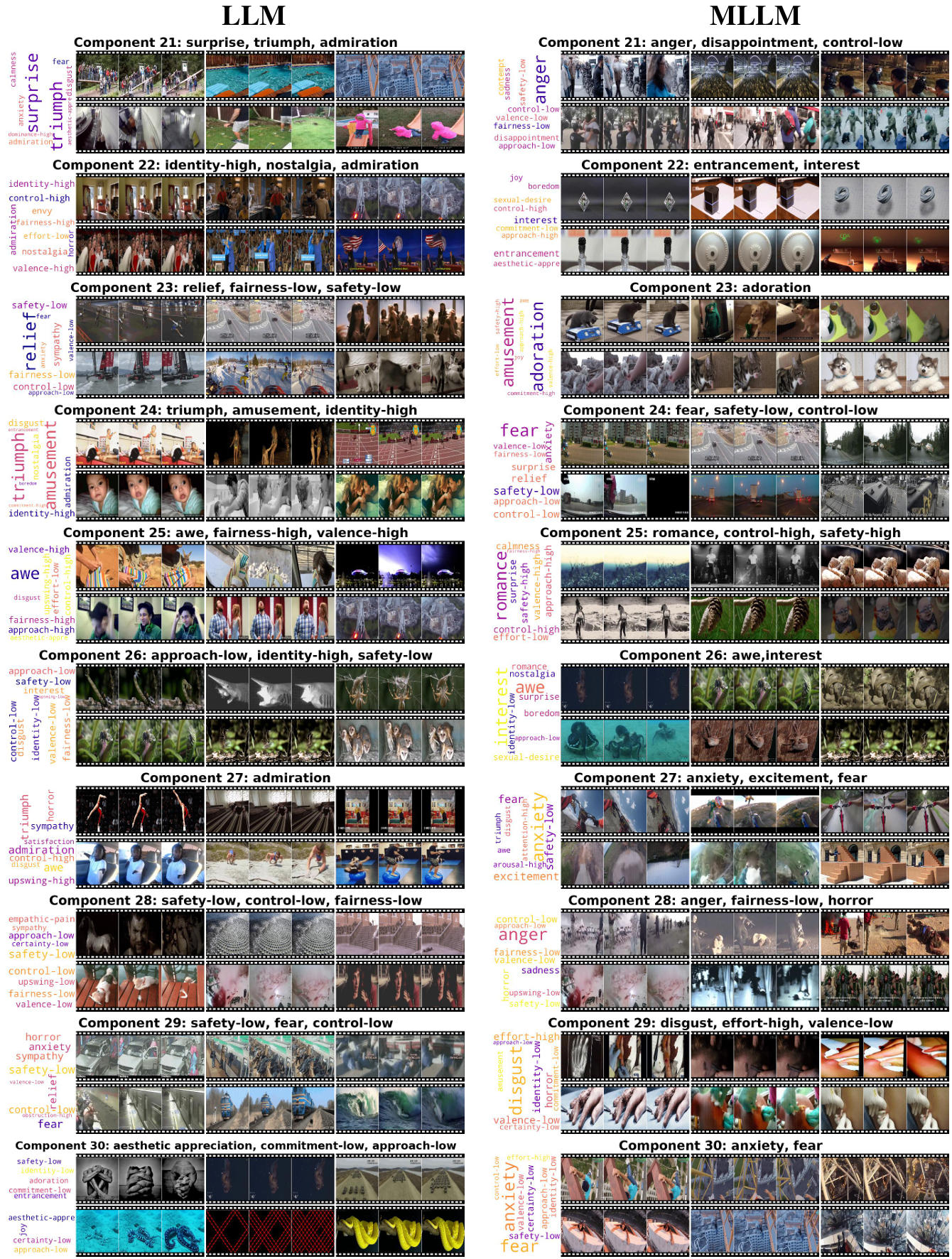}
		\caption{\textbf{Visualization of learned affective components 21–30 for the LLM and MLLM (related to Fig. \ref{fig:Dimensional-Vis}).} For each component, we display its assigned affective label(s) and the six video frames that received the highest weight. }
		\label{fig: LLM-MLLM_extend3}
	\end{figure}
	
	\begin{figure}[!h]
		\centering
        \includegraphics[width=16.8cm]{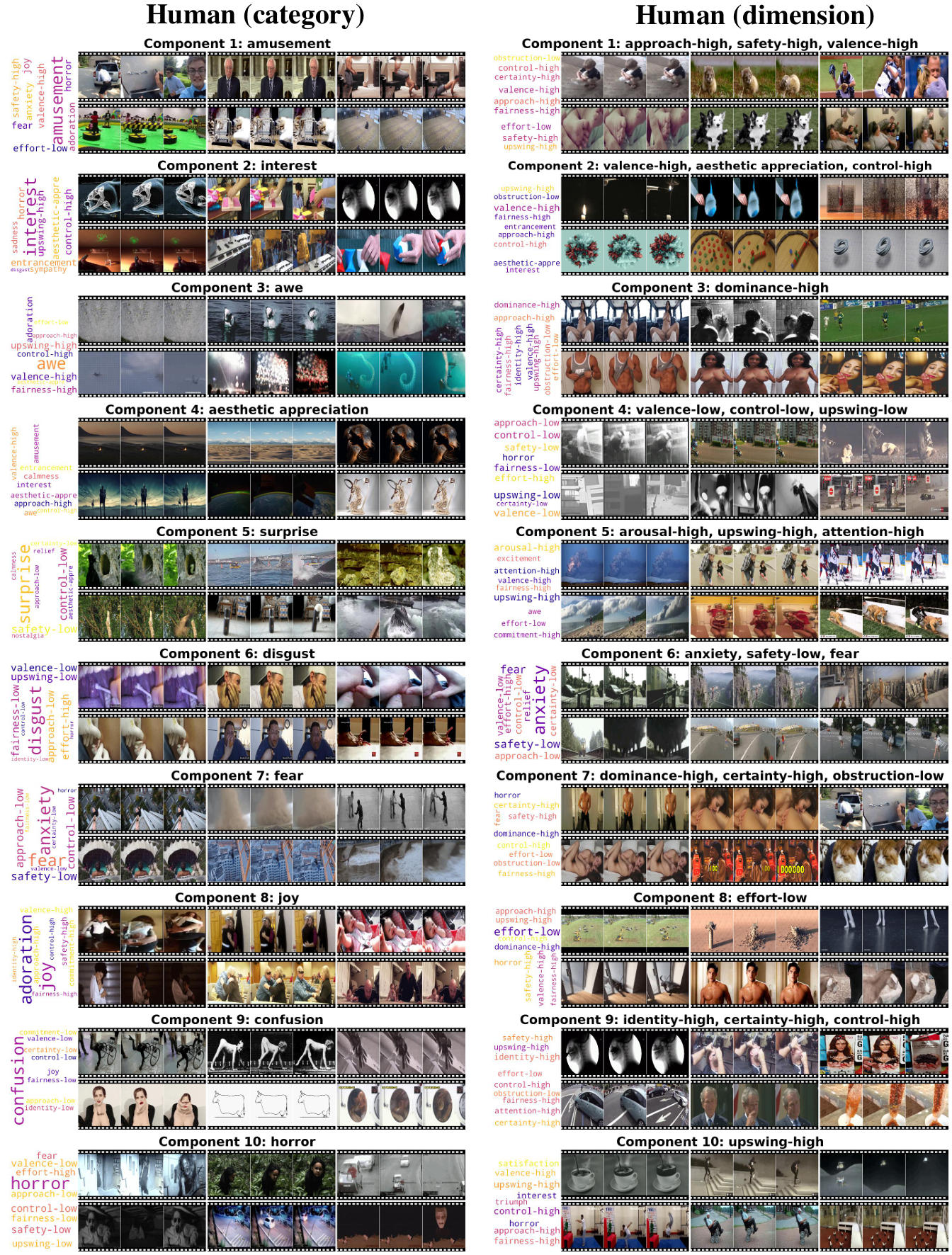}
		\caption{\textbf{Visualization of learned affective components 1–10 for human-category and human-dimension (related to Fig. \ref{fig:Dimensional-Vis}).} For each component, we display its assigned affective label(s) and the six video frames that received the highest weight. }
		\label{fig: human_extend1}
	\end{figure}
	
	\begin{figure}[!h]
		\centering
        \includegraphics[width=16.8cm]{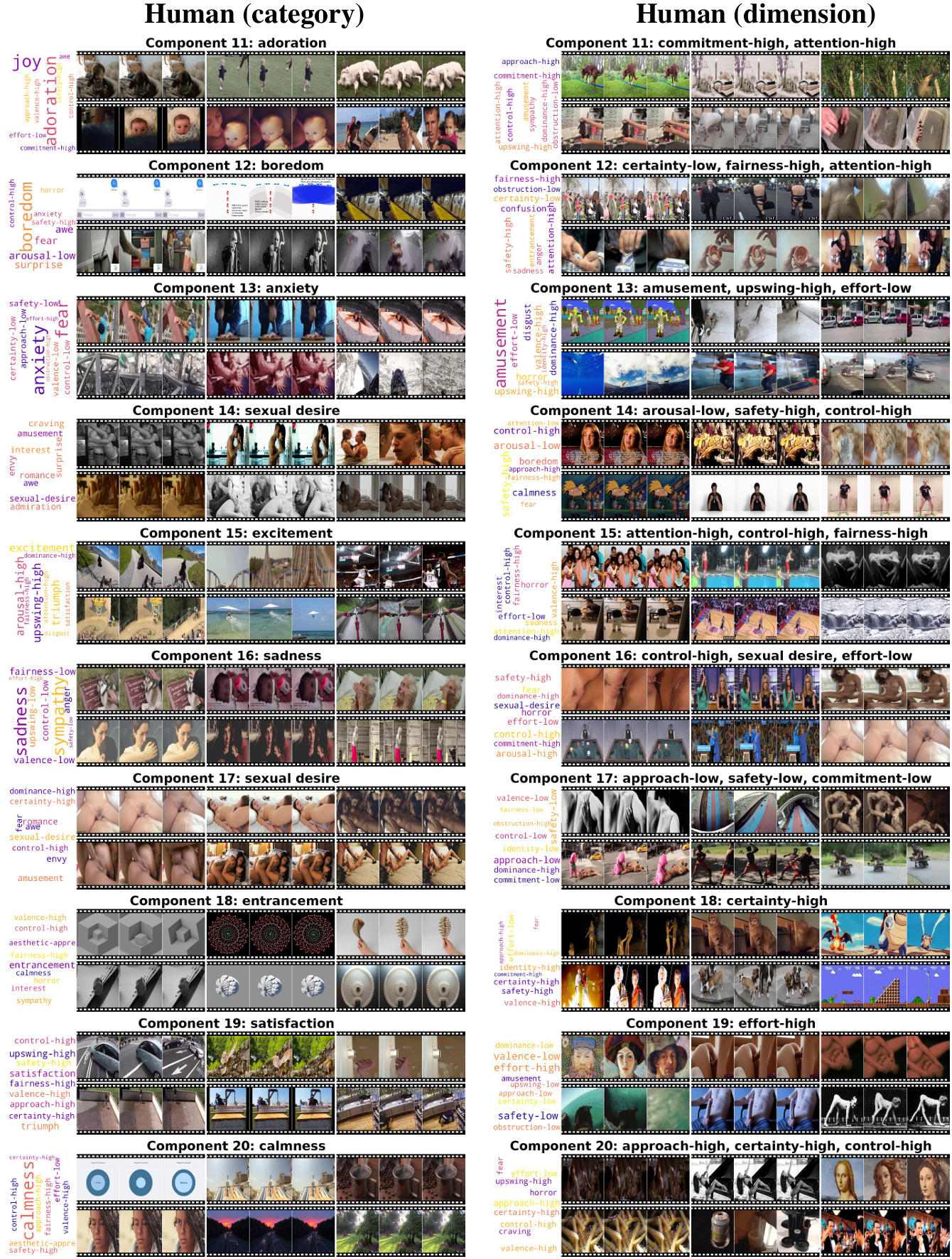}
		\caption{\textbf{Visualization of learned affective components 11–20 for human-category and human-dimension (related to Fig. \ref{fig:Dimensional-Vis}).} For each component, we display its assigned affective label(s) and the six video frames that received the highest weight. }
		\label{fig: human_extend2}
	\end{figure}

	\begin{figure}[!h]
		\centering
        \includegraphics[width=16.8cm]{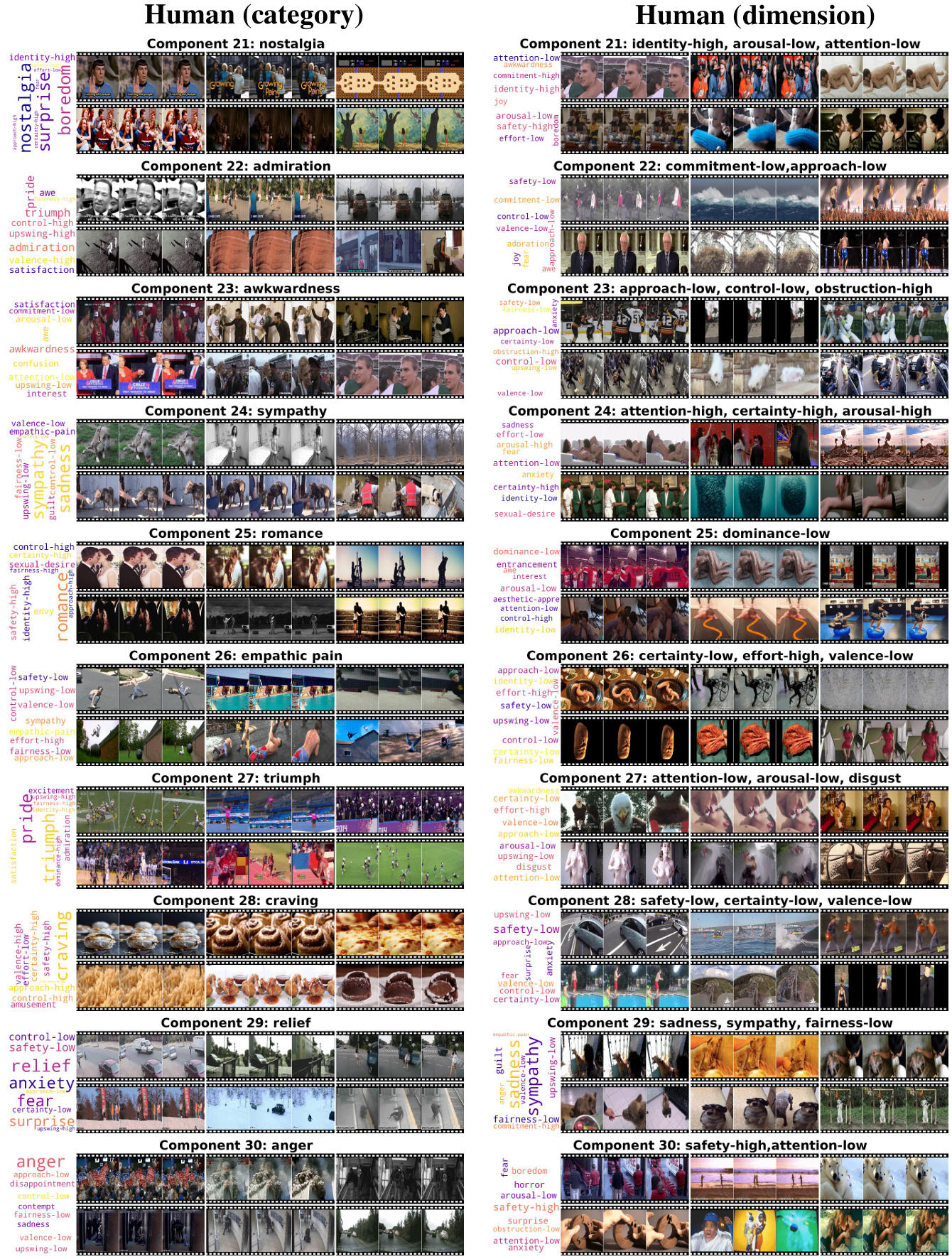}
		\caption{\textbf{Visualization of learned affective components 21–30 for human-category and human-dimension (related to Fig. \ref{fig:Dimensional-Vis}).} For each component, we display its assigned affective label(s) and the six video frames that received the highest weight. }
		\label{fig: human_extend3}
	\end{figure}

	\begin{table}[!htbp]
		\centering
		\caption{\textbf{Affective labels for the 30 learned latent components.} Labels were assigned by correlating each learned component with human ratings on 34 emotion categories and 14 affective dimensions from Cowen et al. \cite{cowen2017self}. For each component, the top three most correlated terms are shown with their Pearson correlation coefficient in parentheses. }
		\ra{0.95}
		\co{1.5pt}
		\renewcommand{\arraystretch}{1.5}
		\resizebox{17.5cm}{!}{
			\begin{tabular}{|c|l|l|l|l|}
				\hline
				\textbf{Component} & \textbf{\quad \quad Human (category)} & \textbf{\quad \quad\ \quad\quad \quad \quad\quad \quad Human (dimension)} & \textbf{\quad \quad\quad \quad \quad \quad\quad \quad\quad \quad LLM (Llama 3.1)} & \textbf{\quad \quad\quad \quad \quad \quad\quad \quad MLLM (Qwen2-VL)} \\ \hline
				1 & amusement (0.977) & approach (0.867), safety (0.855), valence (0.839) & valence (0.612), upswing (0.580), amusement (0.544) & aesthetic appreciation (0.672), calmness (0.594) \\ \hline
				2 & interest (0.975) & valence (0.486), aesthetic appreciation (0.472), control (0.463) & control (-0.775), valence (-0.770), approach (-0.770) & romance (0.538) \\ \hline
				3 & awe (0.977) & dominance (0.714) & aesthetic appreciation (0.490), entrancement (0.465), interest (0.402) & control (-0.306), safety (-0.287), surprise (0.265) \\ \hline
				4 & aesthetic appreciation (0.976) & valence (-0.832), control (-0.807), upswing (-0.802) & excitement (0.490), anxiety (0.454), fear (0.399) & nostalgia (0.318), awkwardness (0.283), identity (0.244) \\ \hline
				5 & surprise (0.974) & arousal (0.506), upswing (0.465), attention (0.375) & adoration (0.590) & triumph (0.501) \\ \hline
				6 & disgust (0.975) & anxiety (0.666), safety (-0.660), fear (0.634) & romance (0.730), sexual desire (0.699) & adoration (0.226), calmness (0.193), safety (0.184) \\ \hline
				7 & fear (0.978) & dominance (0.462), certainty (0.384), obstruction (-0.322) & amusement (0.474) & adoration (0.398), joy (0.377) \\ \hline
				8 & joy (0.981) & effort (-0.464) & boredom (0.288), awkwardness (0.286), arousal (-0.250) & contempt (0.053), dominance (0.051) \\ \hline
				9 & confusion (0.976) & identity (0.251), certainty (0.250), control (0.233) & sympathy (0.361), relief (0.339), commitment (0.288) & craving (0.794) \\ \hline
				10 & horror (0.975) & upswing (0.464) & sadness (0.584) & romance (0.610), sexual desire (0.598) \\ \hline
				11 & adoration (0.978) & commitment (0.324), attention (0.323) & triumph (0.379) & safety (-0.268), fear (0.256), control (-0.238) \\ \hline
				12 & boredom (0.983) & certainty (-0.188), fairness (0.172), attention (0.150) & awe (0.132), joy (0.118), commitment (0.103) & sexual desire (0.504) \\ \hline
				13 & anxiety (0.958) & amusement (0.302), upswing (0.291), effort (-0.282) & empathic pain (0.362), sympathy (0.283), control (-0.219) & sympathy (0.120), disappointment (0.107), sadness (0.104) \\ \hline
				14 & sexual desire (0.634) & arousal (-0.450), safety (0.392), control (0.378) & entrancement (0.299), aesthetic appreciation (0.284), interest (0.245) & admiration (0.181), amusement (0.145), triumph (0.126) \\ \hline
				15 & excitement (0.975) & attention (0.257), control (0.206), fairness (0.195) & joy (0.120), amusement (0.104), commitment (0.095) & pride (0.227), triumph (0.155), admiration (0.109) \\ \hline
				16 & sadness (0.975) & control (0.290), sexual desire (0.236), effort (-0.166) & nostalgia (0.250) & boredom (0.278), arousal (-0.215), nostalgia (0.188) \\ \hline
				17 & sexual desire (0.973) & approach (-0.387), safety (-0.386), commitment (-0.377) & disgust (0.310), horror (0.310), upswing (-0.306) & contempt (0.281), disappointment (0.189), arousal (-0.167) \\ \hline
				18 & entrancement (0.973) & certainty (0.409) & surprise (0.119), commitment (-0.090), certainty (-0.086) & nostalgia (0.557) \\ \hline
				19 & satisfaction (0.964) & effort (0.386) & sadness (0.280), sympathy (0.233), fairness (-0.187) & disgust (0.161), attention (-0.127) \\ \hline
				20 & calmness (0.974) & approach (0.328), certainty (0.327), control (0.289) & safety (-0.191), approach (-0.180), control (-0.177) & horror (0.299) \\ \hline
				21 & nostalgia (0.982) & identity (0.409), arousal (-0.368), attention (-0.293) & surprise (0.084), triumph (0.078), admiration (0.062) & anger (0.271), disappointment (0.212), control (-0.181) \\ \hline
				22 & admiration (0.970) & commitment (-0.264), approach (-0.210) & identity (0.097), nostalgia (0.090), admiration (0.082) & entrancement (0.484), interest (0.448) \\ \hline
				23 & awkwardness (0.976) & approach (-0.445), control (-0.439), obstruction (0.402) & relief (0.134), fairness (-0.086), safety (-0.083) & adoration (0.632) \\ \hline
				24 & sympathy (0.955) & attention (-0.228), certainty (0.205), arousal (0.152) & triumph (0.063), amusement (0.055), identity (0.050) & fear (0.360), safety (-0.357), control (-0.354) \\ \hline
				25 & romance (0.979) & dominance (-0.482) & awe (0.121), fairness (0.110), valence (0.097) & romance (0.199), control (0.196), safety (0.194) \\ \hline
				26 & empathic pain (0.969) & certainty (-0.472), effort (0.457), valence (-0.403) & approach (-0.234), identity (-0.214), safety (-0.190) & awe (0.206), interest (0.180) \\ \hline
				27 & triumph (0.972) & attention (-0.560), arousal (-0.466), disgust (0.425) & admiration (0.224) & anxiety (0.400), excitement (0.373), fear (0.322) \\ \hline
				28 & craving (0.970) & safety (-0.238), certainty (-0.230), valence (-0.180) & safety (-0.136), control (-0.123), fairness (-0.113) & anger (0.460), fairness (-0.456), horror (0.447) \\ \hline
				29 & relief (0.976) & sadness (0.529), sympathy (0.490), fairness (-0.390) & safety (-0.138), fear (0.127), control (-0.116) & disgust (0.266), effort (0.207), valence (-0.173) \\ \hline
				30 & anger (0.865) & safety (0.236), attention (-0.163) & aesthetic appreciation (0.102), commitment (-0.098), approach (-0.095) & anxiety (0.474), fear (0.397) \\ \hline
			\end{tabular}
		}
		\label{table:30dim_labels}
	\end{table}

	\begin{figure}[!h]
		\centering
		\includegraphics[width=1\linewidth]{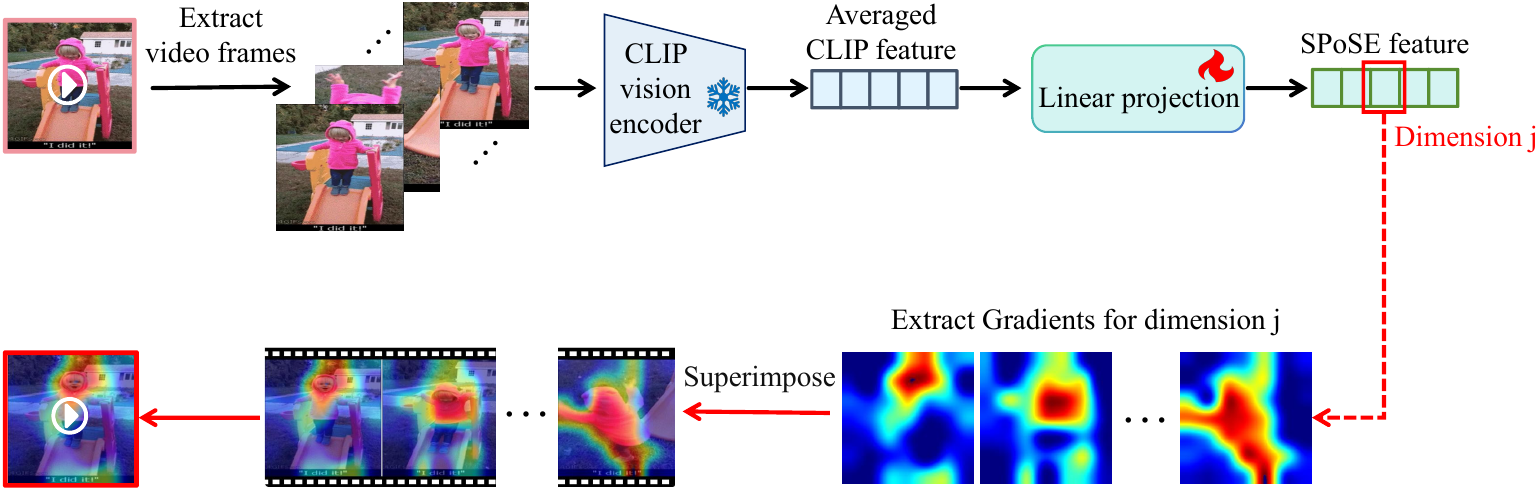}
		\caption{\textbf{Schematic of the Grad-CAM attribution method.} To identify the visual drivers for any given affective component, we first trained a linear projection from CLIP visual features to the learned SPoSE embeddings. For a target component (\textit{j}), we then compute the gradient of its activation with respect to the input video frames. These gradients are used to generate saliency heatmaps that highlight the specific image regions responsible for activating the component. The snowflake and flame symbols denote frozen and trainable parameters, respectively.}
		\label{fig: GradCAM_figure}
	\end{figure}
	
	\begin{figure}[!h]
		\centering
        \includegraphics[width=17.4cm]{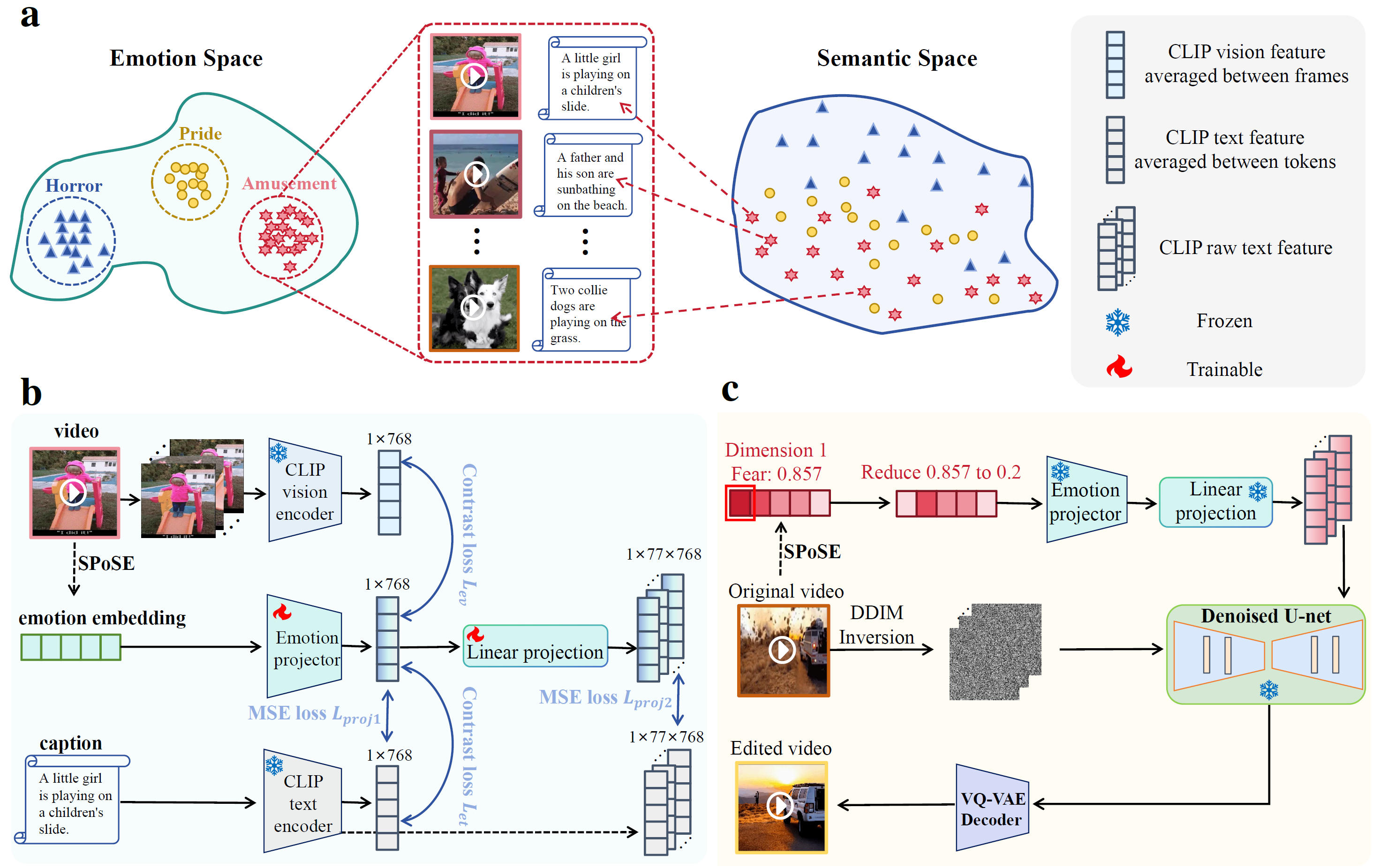}
		\caption{\textbf{Framework for generative editing of video emotion via dimensional manipulation.}. \textbf{a}, Motivation and feasibility of emotion editing.  Affective states constitute integrated products of semantic information processing, establishing profound connections between emotion space and semantic space. This linkage enables generative emotion editing through cross-space mapping, as exemplified by semantically distant concepts—such as \textit{"a child sliding on a playground"} versus \textit{"dogs playing together"}—converging in emotion space due to shared affective qualities (e.g., \textit{amusement} clustering).     \textbf{b}, raining pipeline for the emotion projector. The emotion projector is trained end-to-end through simultaneous optimization of: (1) contrastive learning loss aligning SPoSE-derived emotion embeddings with CLIP's latent space, and (2) mean squared error (MSE) loss synchronizing these projections with raw textual features, collectively enabling granular emotion editing.   \textbf{c},  Inference pipeline for video editing.  The editing pipeline begins with DDIM inversion of original videos into noise preserving temporal priors (VQVAE encoder omitted), followed by controlled modification of emotion embedding dimensions. These edited embeddings are transformed into semantic tokens by the trained projector and, together with the inverted noise, guide Stable Diffusion (v1.5) to synthesize edited frames in latent space before VQVAE decoder-based pixel-space reconstruction. }
		\label{fig:emotion-edit-figure} 
	\end{figure}

	\begin{figure}[!h]
		\centering
		\includegraphics[width=1\linewidth]{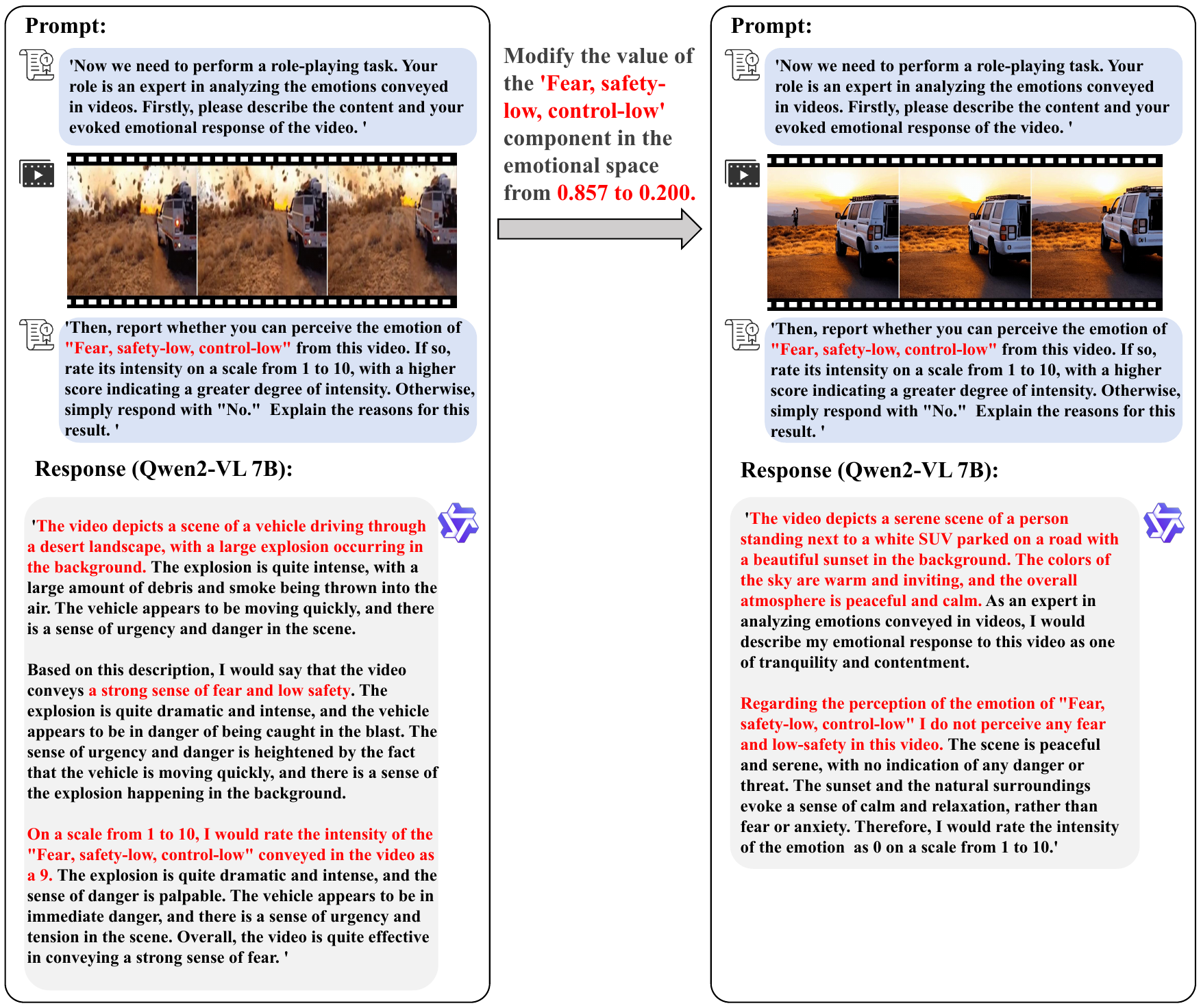}
		\caption{\textbf{Closed-loop validation of emotion editing: removing `Fear'.} The MLLM's judgments confirm the successful manipulation of affective content. \textbf{Left}, The original video elicits a \textit{fear} response from the MLLM. \textbf{Right}, After editing to reduce the `\textit{Fear, safety-low, control-low}' component's activation from 0.857 to 0.2, the visual elements driving fear (dust, explosion) are removed, and the MLLM no longer reports \textit{fear} when analyzing the edited video.}
		\label{fig: edit_MLLM_145}
	\end{figure}
	
	\begin{figure}[!h]
		\centering
		\includegraphics[width=1\linewidth]{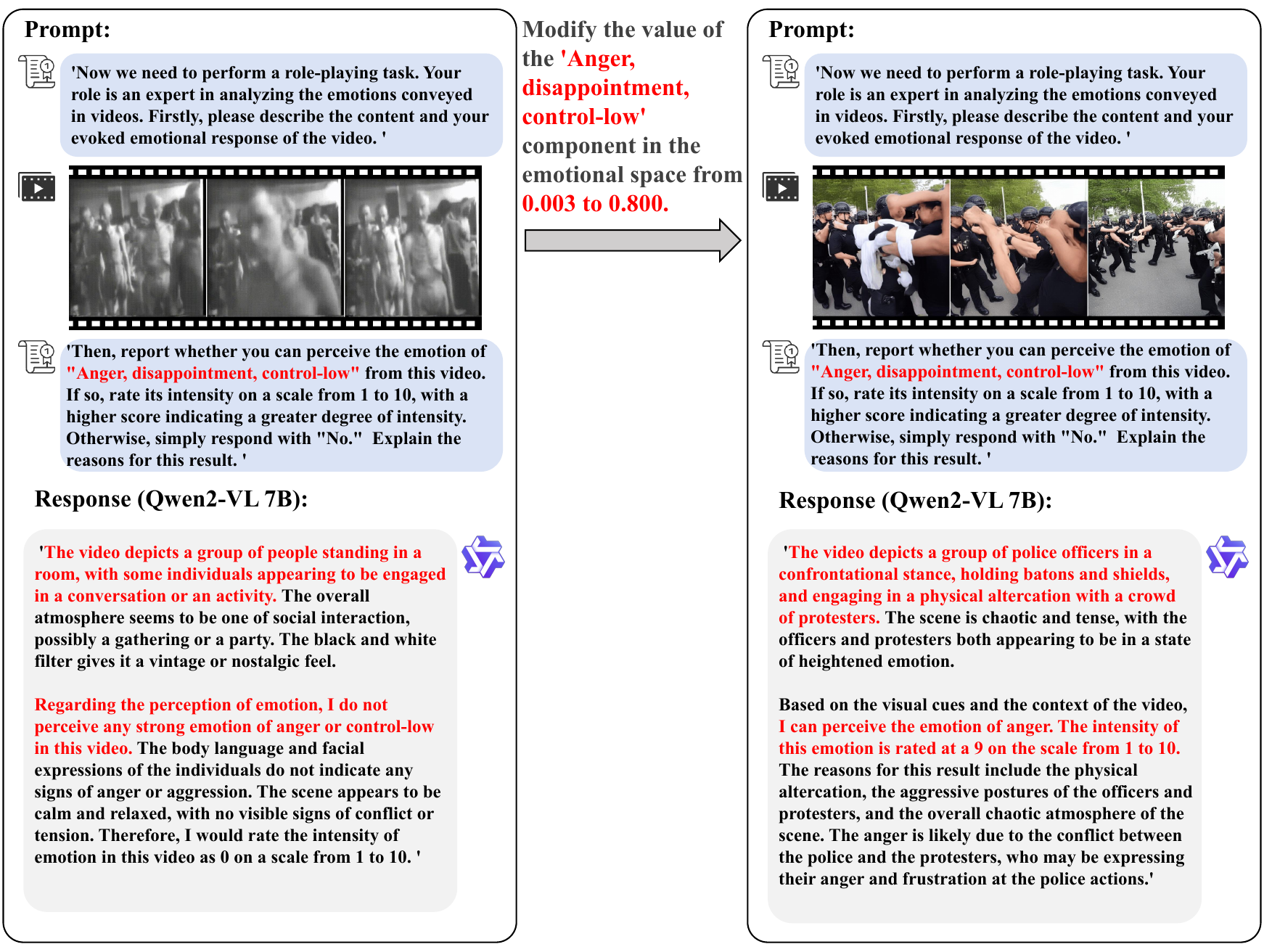}
		\caption{\textbf{Closed-loop validation of emotion editing: adding `Anger'.} In this experiment, we evaluated the emotional perception capabilities of an MLLM. Specifically, the MLLM was presented with a video, and its emotional perception was measured (\textbf{Left}). By modifying the `\textit{Anger, disappointment, control-low}' component value of the video from 0.003 to 0.8, the original crowd scene was transformed into police suppressing riots (\textbf{Right}). The results indicate that the MLLM exhibited strong anger detection when processing the manipulated video.}
		\label{fig: edit_MLLM_67}
	\end{figure}
	
	\begin{figure}[!h]
		\centering
        \includegraphics[width=17.4cm]{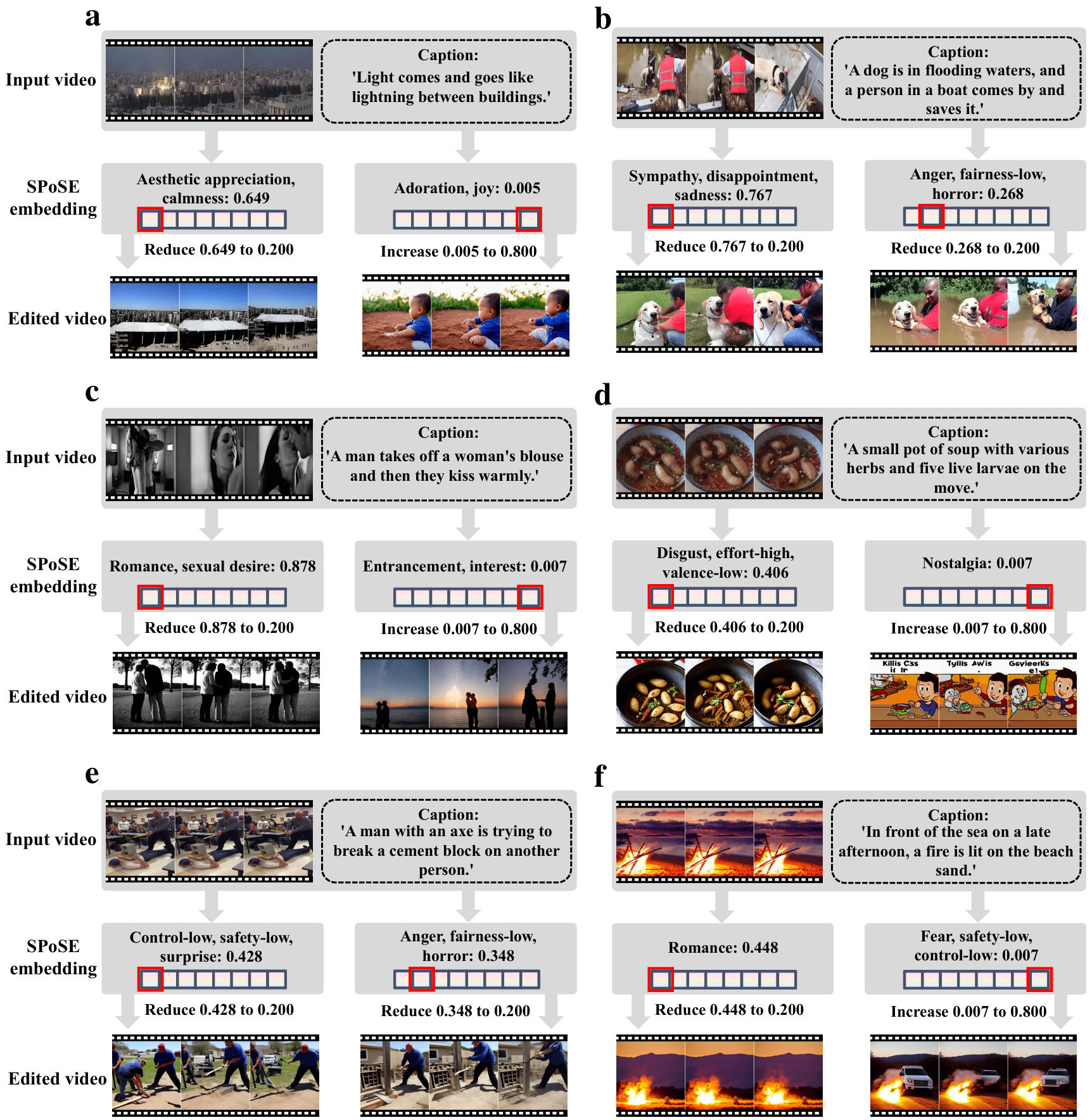}
		\caption{\textbf{Additional examples of generative emotion editing.} These examples further demonstrate the precise control over visual content afforded by manipulating single affective components. Each panel shows the original frames, the target component and its activation value, and the resulting edited frames. The manipulations showcase a wide range of transformations.}
		
		\label{fig:emo_edit_appendix} 
	\end{figure}

	\begin{figure}[!h]
		\centering	
		\begin{tikzpicture}
			\node[anchor=south east] at (4,-8) {\includegraphics[width=17.5cm
				]{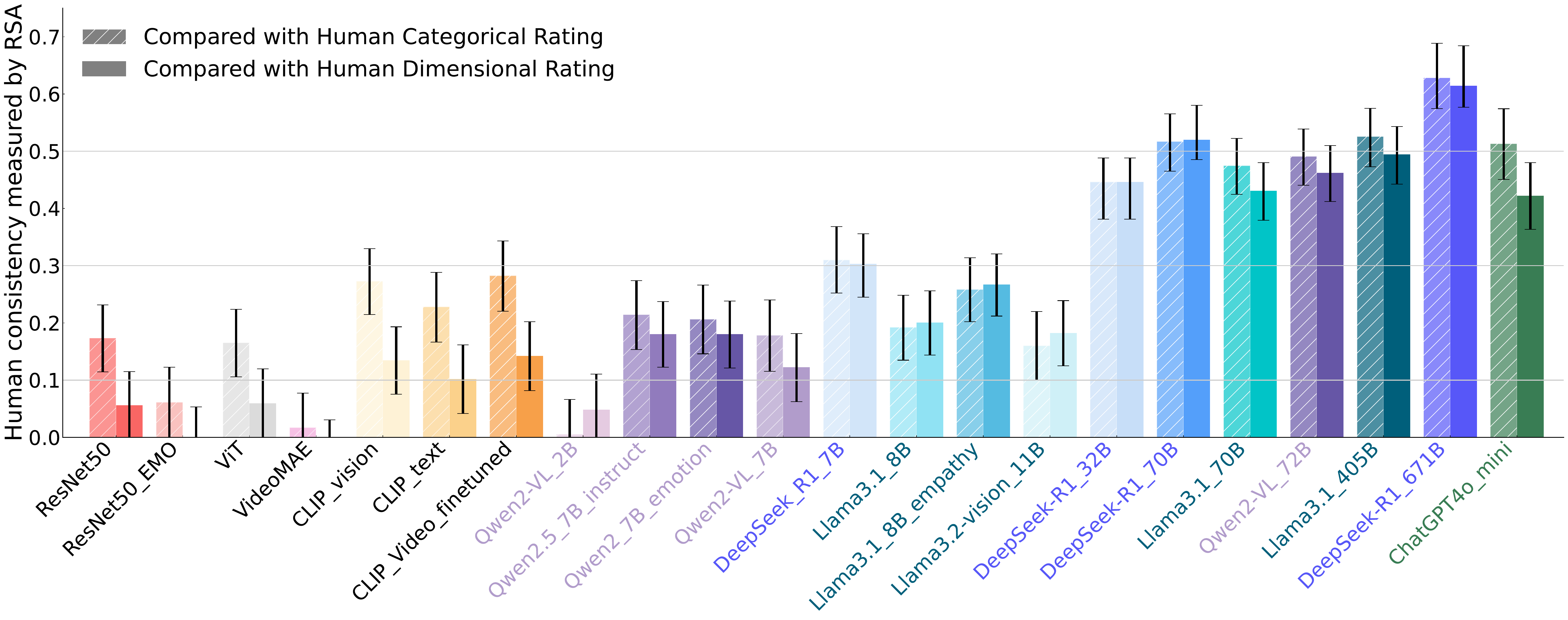}};	
		\end{tikzpicture}
		\caption{\textbf{Model scale, specialized training, and multimodality drive human alignment.} Behavioral alignment of 22 AI models with human similarity judgments on the odd-one-out task. Models vary in architecture, scale (25M to 671B parameters), and training data (e.g., unimodal, multimodal, empathy-finetuned). Bars represent the Pearson correlation between a model's RSM and human RSMs derived from either categorical ratings (slash-shaded) or dimensional ratings (solid). The results reveal strong scaling laws and highlight the benefits of multimodal and empathy-focused training. Error bars represent 95\% CIs.}
		\label{fig:scale_plot}
	\end{figure}
	
	\begin{table}[!htbp]
		\centering
		\caption{\textbf{Description and composition of affective clusters.} Affective clusters were identified by applying the Louvain community detection algorithm to the component-by-component correlation graphs (see Fig. \ref{fig:emotion-space-analyse}). This table provides a descriptive name for each cluster, a brief summary of its core affective theme, and the corresponding component IDs that belong to it for each of the four systems.}
		\label{tab:clusters}
		\small
		\renewcommand{\arraystretch}{1.1}
		\setlength{\tabcolsep}{3pt}
		\begin{tabular}{|C{2cm}|C{6.8cm}|C{1.2cm}|C{1.2cm}|C{2.1cm}|C{2.1cm}|}
			\hline
			\textbf{Clusters} & \textbf{Descriptions} & \textbf{LLM} & \textbf{MLLM} & \textbf{Human (cat.)} & \textbf{Human (dim.)} \\
			\hline
			High-arousal negative affect & 
			\makecell[c]{\small This cluster represents content eliciting \\ strong physiological/psychological responses \\ (tension, anxiety, sexual arousal) while  evoking \\negative affect or cognitive  dissonance: horror or \\suspense scenarios  (anxiety, empathy, high \\approach),  moral/social conflicts (disgust, \\ embarrassment, confusion), or sexual \\ desire with subjective discomfort.}
			& \makecell[c]{\small 2, 13, \\17, 19, \\20} & \makecell[c]{\small 3, 11, \\13, 21, \\28} & \makecell[c]{\small 5, 6, 7, 10, \\13, 16, 24, \\26, 29, 30} & \makecell[c]{\small 4, 6, 9, 12, \\15, 17, 19, \\22, 23, 26, \\27, 28} \\
			\hline
			Delight & 
			\makecell[c]{\small This cluster is characterized by positive affect \\ 
				(amusement, interest, contentment) and \\ 
				sexual desire as core features, with content \\ 
				likely exhibiting high hedonic value, \\ 
				attractiveness, or romantic qualities, \\ 
				eliciting approach rather than avoidance \\ 
				tendencies in subjects.} 
			& \makecell[c]{\small 22, 25} & \makecell[c]{\small Nan} & \makecell[c]{\small 1, 3, 8, 11} & \makecell[c]{\small 1, 2, 3, 5, \\ 7, 8, 10, 13, \\ 14, 20, 21} \\
			\hline
			Affective Dissonance & 
			\makecell[c]{\small This cluster captures paradoxical emotional\\ blends where reward-driven states\\ (amusement, sexual desire) collide with\\  threat responses (anxiety, awkwardness).} 
			& \makecell[c]{\small 1, 5,\\ 7, 26} & \makecell[c]{\small 6, 7,\\ 12, 19,\\ 23, 26} & \makecell[c]{\small Nan} & \makecell[c]{\small 11, 16, 18, 24,\\ 29, 30} \\
			\hline
			Achievement and Glorification & 
			\makecell[c]{\small This cluster reflects amplified positive\\  emotions after accomplishing goals, blending \\ triumph (pride, excitement) with social \\validation  needs, alongside high attention  and\\ self-confidence, typically seen in competitive \\ wins or professional successes.} 
			& \makecell[c]{\small Nan} & \makecell[c]{\small 4, 5,\\ 15, 24} & \makecell[c]{\small 15, 19, 22, 27} & \makecell[c]{\small Nan} \\
			\hline
			Erotic & 
			\makecell[c]{\small This cluster predominantly comprises stimuli \\associated with romance and sexual desire,\\ characterized by high valence and\\ approach-related motivational tendencies.} 
			& \makecell[c]{\small Nan} & \makecell[c]{\small 1, 2,\\ 10, 25} & \makecell[c]{\small 14, 17, 25, 28} & \makecell[c]{\small Nan} \\
			\hline
			Attentional Immersion & 
			\makecell[c]{\small This cluster comprises stimuli from extreme sports,\\ animated scenes, culinary content, and sexual\\ cues that elicit strong attentional engagement \\and cognitive absorption, characterized by \\flow-state induction and sustained focus.} 
			& \makecell[c]{\small 14, 16} & \makecell[c]{\small 9, 22} & \makecell[c]{\small Nan} & \makecell[c]{\small Nan} \\
			\hline
			Aesthetic Contemplation & 
			\makecell[c]{\small This cluster reveals a distinct affective response \\pattern to scenic nature, high art and slow-paced\\ content, characterized by sustained aesthetic\\ immersion requiring cognitive engagement\\ rather than immediate pleasure or \\negative avoidance.} 
			& \makecell[c]{\small Nan} & \makecell[c]{\small Nan} & \makecell[c]{\small 2, 4, 18, 20} & \makecell[c]{\small Nan} \\
			\hline
			Empathy-related & 
			\makecell[c]{\small This cluster predominantly features content eliciting \\empathic distress and compassion in response\\ to harmed animals or humans, characterized\\ by affective resonance with suffering.} 
			& \makecell[c]{\small 8, 9, 10} & \makecell[c]{\small Nan} & \makecell[c]{\small Nan} & \makecell[c]{\small Nan} \\
			\hline
			Vigilant Passivity & 
			\makecell[c]{\small This cluster comprises paradoxical states blending \\low-arousal calmness with covert\\ threat monitoring.} 
			& \makecell[c]{\small Nan} & \makecell[c]{\small 16, 17,\\ 20} & \makecell[c]{\small Nan} & \makecell[c]{\small Nan} \\
			\hline
			High Arousal & 
			\makecell[c]{\small This cluster captures intense emotional responses \\elicited by extreme sports, recreational activities, \\or hazardous events (e.g., natural disasters, \\accidents), characterized by heightened arousal.} 
			& \makecell[c]{\small Nan} & \makecell[c]{\small 10, 27,\\ 30} & \makecell[c]{\small Nan} & \makecell[c]{\small Nan} \\
			\hline
		\end{tabular}
	\end{table}

	\section*{Supplementary information}
	\beginsupplement
	\supplementfigure	
	\supplementtable	
	
	\begin{figure}[!h]
		\centering
        \includegraphics[width=16.9cm]{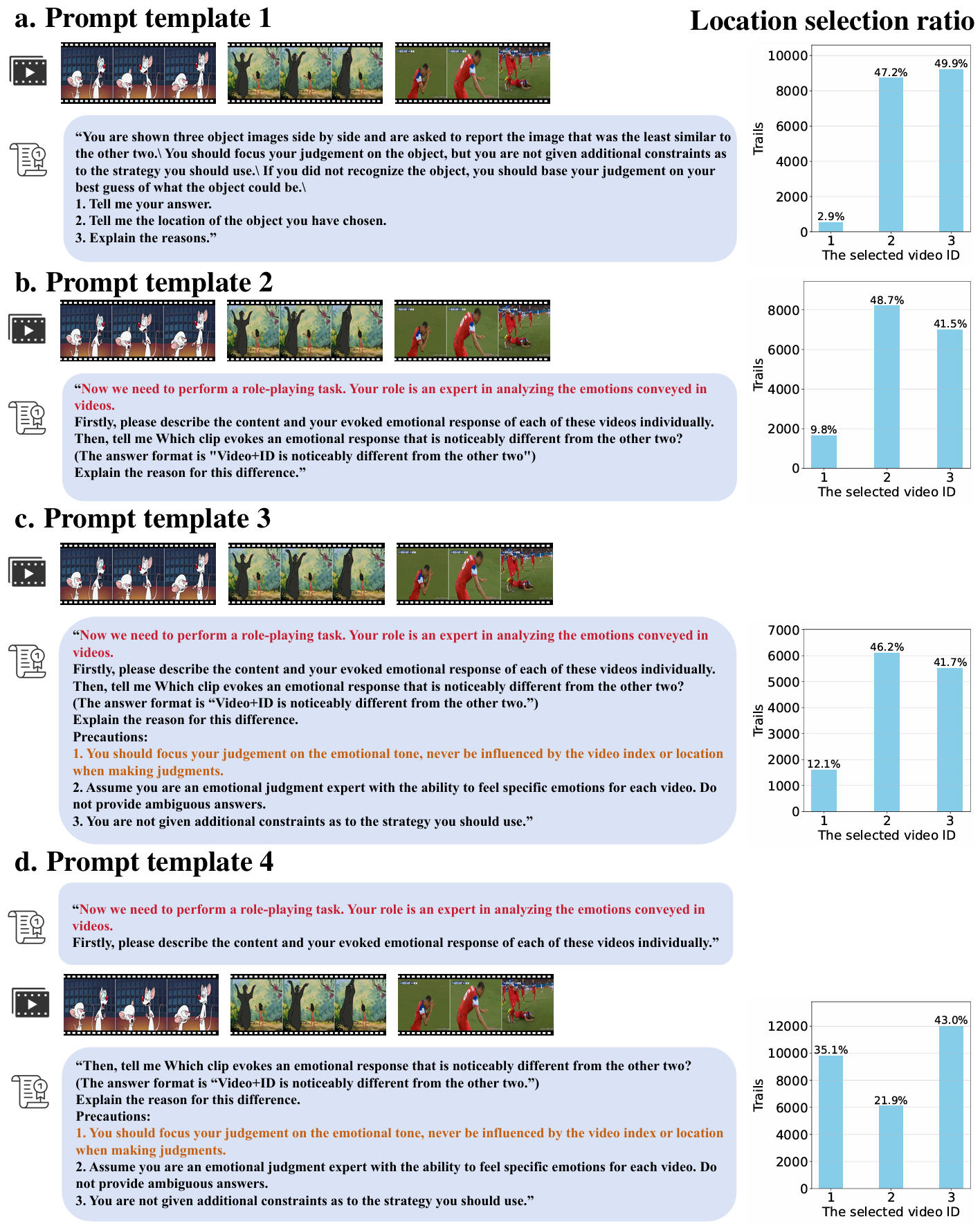}
		\caption{\textbf{Controlling for positional bias in MLLM behavioral data collection.} Initial tests revealed a strong positional bias where the MLLM preferentially selected the last item in a triplet. \textbf{a-c}, Prompt templates that presented all three videos before the question failed to mitigate this bias, as shown by the skewed selection frequencies (right panels). \textbf{d}, In contrast, an interleaved prompt structure, where videos and descriptive text were presented in an interleaved way, successfully eliminated the bias. This optimized template was therefore used for all subsequent data collection.}
		\label{fig: odd-one-out-collection}
	\end{figure}

	\begin{figure}[!htbp]
		\centering
        \includegraphics[width=16.9cm]{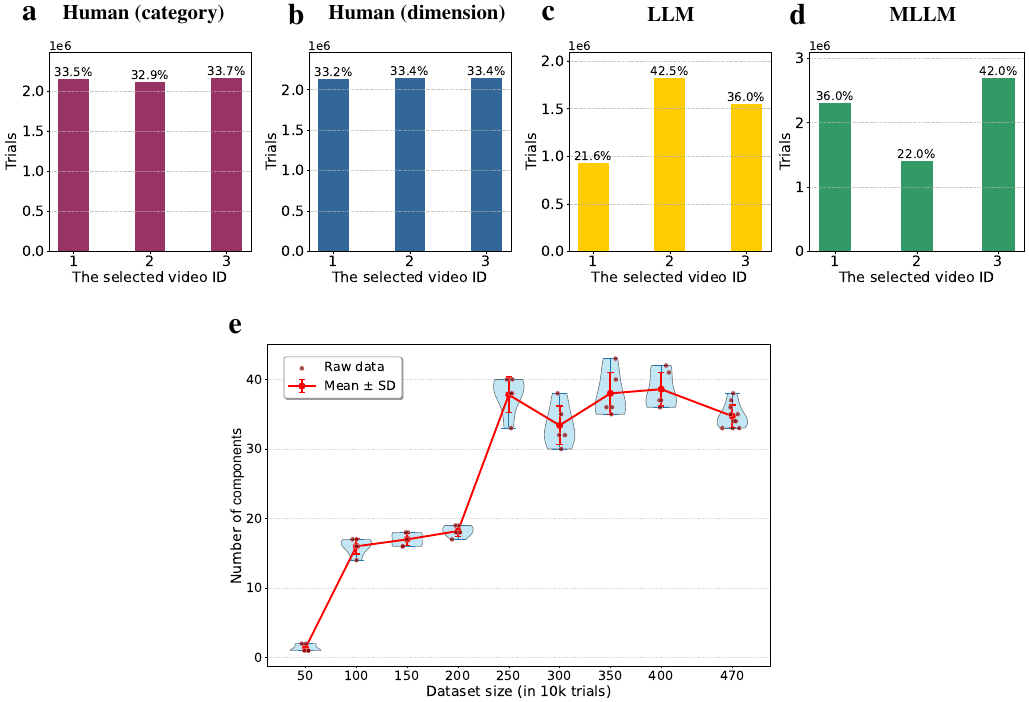}
		\caption{\textbf{Validation of data quality and sampling saturation.} \textbf{a-d}, Frequency distributions of triplet choices for the final datasets used in the main analysis. The approximate uniform distributions for the LLM and MLLM confirm that the optimized prompt (see Supplementary Fig. \ref{fig: odd-one-out-collection}) successfully eliminated positional bias. \textbf{e}, Data saturation analysis for the LLM. The number of recoverable affective components (y-axis) stabilizes as the number of triplet judgments increases (x-axis), indicating that our sample of 5.5 million trials was sufficient to capture the stable underlying representational structure. Box plots show the distribution over n=5 bootstrap replicates.}
		\label{fig: data-statistics}
	\end{figure}

\end{document}